\documentclass[a4paper,12pt]{article}
\pdfoutput=1 
\usepackage{jheppub} 
\usepackage[T1]{fontenc} 
\usepackage{color}

\newcommand{\be}{\begin{equation}}
\newcommand{\ee}{\end{equation}}
\newcommand{\bea}{\begin{eqnarray}}
\newcommand{\eea}{\end{eqnarray}}
\newcommand{\beal}{\begin{aligned}}
\newcommand{\eeal}{\end{aligned}}

\def\mathbi#1{\textbf{\em #1}}

\def\quarter{\textstyle{\frac{1}{4}}}

\title{Vortex hair on AdS black holes}

\author[a,b]{Ruth Gregory,} 
\author[c]{Peter C.~Gustainis,}
\author[b,c]{David Kubiz{\v n}\'ak,}
\author[b,c]{\\Robert B.~Mann,} 
\author[a]{and Danielle Wills}

\affiliation[a]{Centre for Particle Theory, South Road, Durham, DH1 3LE, UK}
\affiliation[b]{Perimeter Institute, 31 Caroline Street North, Waterloo, 
ON, N2L 2Y5, Canada}
\affiliation[c]{Department of Physics and Astronomy, University of Waterloo, Waterloo, Ontario N2L 3G1, Canada}

\emailAdd{r.a.w.gregory@durham.ac.uk}
\emailAdd{pgustain@uwaterloo.ca}
\emailAdd{dkubiznak@perimeterinstitute.ca}
\emailAdd{rbmann@uwaterloo.ca} 
\emailAdd{d.e.wills@durham.ac.uk}

\abstract{
We analyse vortex hair for charged rotating asymptotically AdS black 
holes in the abelian Higgs model. We give analytical and numerical 
arguments to show how the vortex interacts with the horizon of the
black hole, and how the solution extends to the boundary. The 
solution is very close to the corresponding asymptotically flat vortex,
once one transforms to a frame that is non-rotating at the boundary.
We show that there is a Meissner effect for extremal black holes, with
the vortex flux being expelled from sufficiently small black holes. The
phase transition is shown to be first order in the presence of rotation,
but second order without rotation.  We comment on applications to
holography.
}

\keywords{Cosmic strings, Black holes, No hair theorems}
\preprint{DCPT-14/23}

\begin{document}

\maketitle

\section{Introduction}

That black holes have no hair is a long-standing dictum of classical general
relativity \cite{nohair}, one whose content is highly contingent upon 
assumed conditions.  Although the original no-hair theorems were more
about limiting charges a black hole could carry, they have come to be
taken more widely as meaning {\it black holes cannot support nontrivial
fields on their event horizon}. This outlook is supported by the original
no hair theorems for gauge fields and scalars \cite{ChaseAdler}, which
placed what were regarded as eminently reasonable conditions on 
matter fields. In the intervening years, however, it has become clear that 
these conditions are not only too restrictive \cite{nohair2}, but in fact there
are many situations of physical interest in which black holes can 
support nontrivial field configurations. Most of these are concerned 
with asymptotically flat space times \cite{otherhair} whose hair
falls off sufficiently rapidly at large distances from the black hole, though
there are examples of nonsingular cosmological solutions with time
dependence \cite{coshair}, or indeed scalar condensates around
Kerr black holes \cite{Kerrscalar}.

Topological defects form an interesting class of alternative examples 
of black hole hair outside of the asymptotically flat class. 
Both domain walls and cosmic strings \cite{Vilenkin}, topologically 
stable objects with a nontrivial quantum-field-theoretic vacuum 
structure, can have significant gravitational influence, and were   
originally expected to be antipathetic to black holes,
in part because of the problem of how
to have the associated fields end on the event horizon, but
also because of the strong global gravitational impact of the black
hole. Domain walls provide a  `mirror' to spacetime (effectively compactifying
space \cite{IpS}) and cosmic strings yield a conical deficit that 
generates a gravitational lens \cite{Vilenkin2}.
It is now known that  both can ``pierce'' the black hole \cite{EGS,AGK}:
in the former case, the field theoretic wall provides a smooth transition 
between mirror images of the northern hemisphere of the 
C-metric\footnote{An accelerating black hole metric \cite{KinW}.},
whereas in the latter case a smooth version of the Aryal-Ford-Vilenkin
metric \cite{AFV} represents a black hole with a conical 
deficit through its poles. The original solution \cite{AGK} has 
been generalized in a number of ways to include vortices ending 
on black holes \cite{RGMH}, charged black holes 
\cite{CCESa,BEG}, dilatonic black holes \cite{ROG},
rotating black holes \cite{GB,GKW}, and asymptotically 
dS \cite{GMdS} and AdS black holes \cite{VH,DGM1}.
Fields typically terminate on the event horizon or, in the case of 
extremal black holes, be expelled from the horizon if the width 
of the string is comparable to the size of the black hole. 

Most recently, the rotating black hole has been subject to a thorough
study \cite{GKW}, which analysis corrected earlier work
that had a flawed ansatz \cite{GB}. There is now a detailed understanding 
of how the core fields of a vortex accommodate the rotation of asymptotically 
flat black holes and their  associated `electric' field generation. 
The vortex cuts out a local co-rotating deficit azimuthal angle, which leads 
to some novel features, shifting  the ergosphere of the black hole and altering
the innermost stable circular orbit (ISCO). As with charged black holes, 
flux expulsion  can indeed take place under certain circumstances.  
However unlike the charged case the phase transition is of first order 
and numerical evidence suggests that the flux-expelled solution is 
not dynamically stable.

Here, we investigate the impact of a  negative cosmological constant 
on the problem of a vortex piercing a black hole.  Specifically, we 
obtain vortex solutions for an Abelian Higgs model minimally coupled 
to Einstein gravity in four dimensions with a negative cosmological 
constant. We obtain both approximate and numerical vortex solutions 
to  the field equations of the Abelian Higgs model in the background 
of a Kerr-Newman-AdS black hole.  We find that as the AdS length,
$\ell$, becomes comparable to the size of the vortex, the core of the 
vortex increasingly narrows and the fields exhibit asymptotic  power-law 
falloff instead of exponential. We find that the Meissner effect, observed  
previously for extremal  Kerr and Reissner-Nordstrom black holes, 
persists here as well, and is first
order if there is non-zero rotation but is otherwise 2nd order.
We find that the flux can pierce the horizon provided
the AdS length is sufficiently large, and numerically obtain the 
critical radius for the transition from piercing to expulsion.

Our work may have interesting astrophysical implications.  
It has long been known \cite{Wald,Expulsion1} that 
spinning black holes tend to expel magnetic fields in a continuous 
way as the black hole is spun up.  Indeed, it has
been argued that all stationary, axisymmetric magnetic fields are 
expelled from the Kerr horizon in the extremal limit
\cite{Expulsion2}.  Since  a Killing vector in the vacuum 
spacetime can act as a vector potential for a Maxwell test field, 
as the hole is `spun up' toward extremality, the component of 
the magnetic field  normal to the horizon approaches zero,
and so the  flux lines are expelled (a phenomenon that also 
occurs for black strings and $p$-branes \cite{Expulsion3}).
This Meissner-like effect could quench the  power of astrophysical 
jets, since the magnetic fields need to pierce the horizon to extract 
rotational energy from the black hole, though it has been recently 
argued \cite{Penna} that split-monopole magnetic fields may 
continue to power black hole jets, with the fields becoming 
entirely radial near the horizon, avoiding expulsion.   
In contrast to this we find (as for the asymptotically flat case 
\cite{GKW}) in the Abelian Higgs model that for large AdS 
black holes the vortex pierces  the event horizon, whereas 
flux is expelled if the black hole is sufficiently small. 
 
From a holographic perspective, a vortex in the bulk has an 
interpretation as a defect in the the dual CFT \cite{VH,Dias:2013bwa}, 
corresponding  in the dual superfluid to  heavy pointlike excitations 
around  which the phase of the condensate winds.  
We comment briefly at the end of our paper on a holographic 
interpretation of our results.

\section{Abelian Higgs model for a cosmic string}

The abelian Higgs model is the canonical toy model for a cosmic
string, as it has the simplest action with the requisite vacuum structure
to allow a vortex to form. We write the action as\footnote{We use 
units in which $\hbar=c=1$ and a mostly minus signature.}
\be \label{abhact}
S = \int d^4x \sqrt{-g} \left [ D_{\mu}\Phi ^{\dagger}D^{\mu}\Phi -
{\quarter} {\tilde F}_{\mu \nu}{\tilde F}^{\mu \nu} - {\quarter}\lambda
(\Phi ^{\dagger} \Phi - \eta ^2)^2 \right ]\,,
\ee
where $\Phi$ is the Higgs field, and $A_\mu$ the U(1) gauge
boson with field strength ${\tilde F}_{\mu\nu}$.
As per usual, we rewrite the field content as:
\bea
\Phi (x^{\alpha}) &=& \eta  X (x^{\alpha}) e^{i\chi(x^{\alpha}) }\,,  \\
A_{\mu} (x^{\alpha}) &=& \frac{1}{e} \left [ P_{\mu} (x^{\alpha}) -
\nabla_{\mu} \chi (x^{\alpha}) \right ]\,.
\eea
These fields extract the physical degrees of freedom of the broken symmetric
phase, with $X$ representing the residual massive Higgs field, and $P_\mu$ the
massive vector boson. The gauge degree of freedom,
$\chi$, is explicitly subtracted, although any non-integrable phase 
factors have a physical interpretation as a vortex.

In terms of these new variables, the equations of motion are
\bea \label{vorteqn}
\nabla _{\mu}\nabla ^{\mu} X - P_{\mu}P^{\mu}X + \frac{\lambda\eta^2}{2}
X(X^2 -1) &=& 0\,, \\
\nabla _{\mu}F^{\mu \nu} + 2e^2\eta^2 X^2 P^{\nu}&=& 0\,.
\eea
Because we have not set $G\equiv 1$, we still have the freedom to
fix the units of energy, or $\eta$. We therefore choose to set
$\sqrt{\lambda}\eta=1$, effectively stating our Higgs field has order 
unity mass.
For further use we also introduce the Bogomol'nyi parameter
\cite{Bog}:
\be
\beta = \lambda/2e^2\,,
\ee
indicating the gauge field has mass of order $1/\sqrt{\beta}$.
Alternately, we can rescale the dimensionful parameters $t$ and $r$ 
in the equations of motion: $t\to\sqrt{\lambda}\eta t$, etc.\ and
their corresponding gauge field components $P_t\to P_t/
\sqrt{\lambda}\eta$ -- note $P_\phi$ remains unrescaled however.

A straight static vortex solution will then have the Higgs profile, 
$X_{NO}$, dependent on a single radial variable, $R$ say, 
and the gauge field will have a single angular component, 
$P_\phi = P_{NO}(R)$, where in flat spacetime $X_{NO}$ 
and $P_{NO}$ satisfy the Nielsen-Olesen equations \cite{NO}
\be
\begin{aligned}
X_{NO}'' + \frac{X_{NO}'}{R} &= \frac{P_{NO}^2X_{NO}}{R^2} 
+ \frac12 X_{NO}(X_{NO}^2-1)\,, \\
P_{NO}'' - \frac{P_{NO}'}{R} &= \frac{X_{NO}^2P_{NO}}{\beta}\,.
\end{aligned}
\ee
The profiles of the $X_{NO}$ and $P_{NO}$ fields are highly localized
around $R=0$, and represent a Higgs core in which the U(1)
symmetry is restored with (in this case) a unit of magnetic flux 
threading through. Higher winding strings can be obtained by
replacing $P_{NO} \to NP_{NO}$, although
these are unstable to splitting for $\beta>1$.

Since we are interested in vortices in an anti-de Sitter black
hole background, for future reference we now discuss the vortex
solution in the pure AdS geometry:
\be
\begin{aligned}
ds^2 &= \Bigl(1+\frac{r^2}{\ell^2}\Bigr)dt^2
-\frac{dr^2}{\left(1+\frac{r^2}{\ell^2}\right )}
-r^2 d\theta^2-r^2 \sin^2\!\theta d\phi^2 \\
&=\frac{\ell^2 + R^2}{\ell^2(1-Z^2)} dt^2 
- \frac{\ell^2+R^2}{(1-Z^2)^2}dZ^2
-\frac{\ell^2 dR^2}{\ell^2+R^2} 
-R^2 d\phi^2  \,.
\end{aligned}
\ee
By writing the AdS metric in this second, cylindrical, form we can
see that if we align the vortex in the $\{R,\phi\}$ plane, the equations
of motion will be independent of $Z$, and hence our vortex can once
again be represented by a set of ordinary differential equations:
\be
\begin{aligned}
\left ( 1+\frac{R^2}{\ell^2}\right)P_{0}''
+\left(\frac{2R^2}{\ell^2}-1\right)\frac{P_{0}'}{R}
&= \frac{X_{0}^2P_{0}}{\beta}\,,\\
\left(1+\frac{R^2}{\ell^2}\right)X_{0}''
+\left(\frac{4R}{\ell^2}+\frac{1}{R}\right)X_{0}'
-\frac{P_{0}^2X_{0}}{R^2} -\frac{1}{2}X_{0}(X_{0}^2-1)&=0\,.
\end{aligned}
\label{AdSNO}
\ee
As $R\to0$, the additional terms dependent on the AdS background
drop away, and we have a very similar field structure on axis to the
Nielsen-Olesen vortex. For $R\gtrsim\ell$ however, the functions are
modified, and the asymptotic fall-off of the fields becomes power
law rather than exponential.
\begin{figure}
\begin{center}
\includegraphics[width=0.9\textwidth]{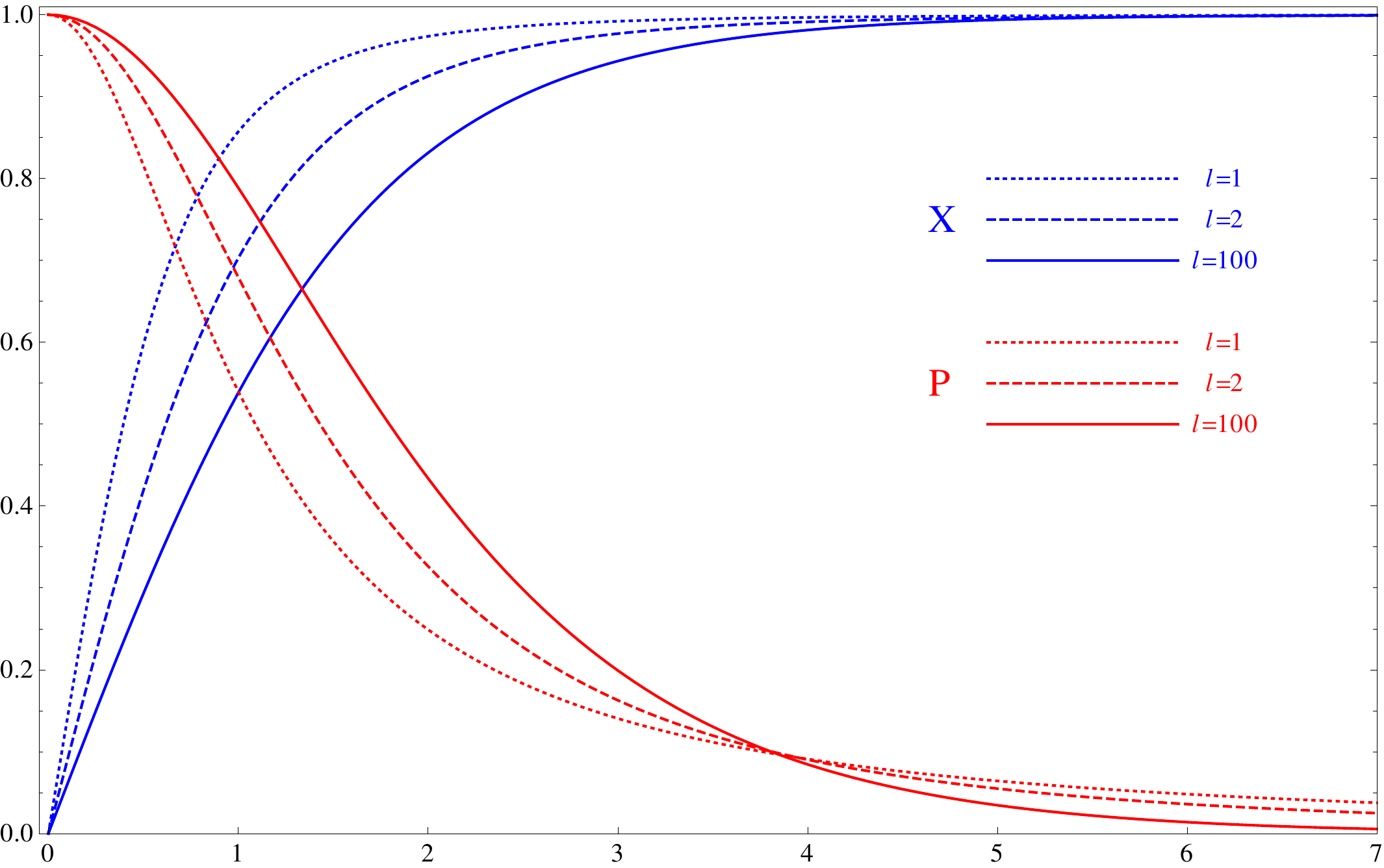}
\caption{{\bf AdS-NO vortex:}
The values of $X$ and $P$ for the AdS NO vortex are 
depicted as functions of $R$.}
\label{fig:adsNO}
\end{center}
\end{figure}

In figure \ref{fig:adsNO} we show the Higgs and gauge profiles for
the AdS vortex. At large $\ell$, the profile is essentially the same as
the pure NO-vortex. However as $\ell$ approaches the scale of the
vortex, the core is seen to narrow, and the power law fall-off becomes
more apparent. Although we can formally integrate these equations
for $\ell \lesssim 1$, it is unclear that such solutions with our boundary
conditions are physically
relevant, as the false vacuum $X=0$ becomes stable for Compton
wavelengths above the AdS scale \cite{BF}.
 
%%%%%

\section{Vortices in Kerr-AdS: Analytics}

Although the full exact solution of a vortex in a black hole background
must be found numerically, there are two ways in which
we can gain insight into the system analytically. The first is by
construction of an approximate solution, and the second is the
case of extremal black holes in which we can prove the existence
(or not) of a piercing solution on the event horizon.

We start by writing down the charged rotating black hole solution
\cite{CarterCMP}
\be
ds^2=\frac{\Delta}{\Sigma}\left[dt-\frac{a\sin^2\!\theta}{\Xi}d\phi\right]^2
-\frac{\Sigma}{\Delta} dr^2-\frac{\Sigma}{S}d\theta^2
-\frac{S\sin^2\!\theta}{\Sigma}\left[a dt-\frac{r^2+a^2}{\Xi}d\phi\right]^2\,,
\label{KNADS}
\ee
where
\bea
\Sigma&=&r^2+a^2\cos^2\!\theta\,,\quad \Xi=1-\frac{a^2}{\ell^2}\,,
\quad S=1-\frac{a^2}{\ell^2}\cos^2\!\theta\,,\nonumber\\
\Delta&=&(r^2+a^2)\Bigl(1+\frac{r^2}{\ell^2}\Bigr)-2mr+q^2\,,
\eea
and the $U(1)$ potential is
\be
A=-\frac{qr}{\Sigma}\left(dt-\frac{a\sin^2\!\theta}{\Xi}d\phi\right)\,.
\ee
The mass $M$, the charge $\mathcal{Q}$, and the angular momentum $J$ 
are related to the parameters $m$, $q$, and $a$ as follows:
\be\label{physical}
GM=\frac{m}{\Xi^2}\,,\quad G\mathcal{Q}
=\frac{q}{\Xi}\,,\quad GJ=\frac{am}{\Xi^2}\,.
\ee
The ergosphere is located at $\Delta = a^2 S \sin^2\theta$,
and the horizon at $\Delta=0$. For large $\ell$, the horizon
is just slightly perturbed from its Kerr-Newman value. As 
$\ell$ decreases, the horizon radius drops, and for small $\ell$
asymptotes to $m^{1/3} \ell^{2/3}$ (or $\sqrt{q\ell}$ for nonzero charge).
We see therefore that for smaller values of $\ell$, the fact that $m\gg1$
is no guarantee that the horizon radius must also be similarly
large in general. However, as we have already remarked, we do
not expect $\ell\lesssim1$ to be physically relevant. Therefore
in any analytic approximation, we will assume $\ell>1$.

Before moving to the vortex equations and analytic results,
it is worth remarking on the behaviour of the horizon radius
in a little more detail, and how this depends on $\ell$. This
is most succinctly captured by the extremal horizon radius,
when $\Delta = \Delta'=0$, which implies
\be
r_+ = \frac{\ell}{\sqrt{6}} \left [ 
\left ( \left ( 1 + \frac{a^2}{\ell^2} \right )^2 
+ 12 \left (\frac{a^2+q^2}{\ell^2}  \right)
\right)^{1/2} - \left ( 1 + \frac{a^2}{\ell^2} \right ) \right] ^{1/2}\,.
\label{rplusext}
\ee
We see therefore that $r_+(a,q,\ell) < \sqrt{a^2+q^2}$, the 
Kerr-Newman value. Moreover, as $\ell$ drops, it is easy to
see that $r_+$ also drops, and for $\ell \lesssim 10$ 
drops quite sharply. Therefore, for the purposes of finding
an approximate solution for the vortex functions, which
typically assumes the black hole is large, we must consider 
$\ell \gtrsim 10$, and for considerations of flux expulsion, 
which typically happens for small black holes, we would expect
any argument to be sensitive to the value of $\ell$.

To find the vortex equations, we must consider not only the $X$
and $P_\phi$ functions, but also a nonzero $P_t$:
\bea
0&=&\Delta X_{,rr}+\Delta' X_{,r}+S X_{,\theta\theta}
+\cot\theta\Bigl(S +\frac{2a^2}{\ell^2}\sin^2\theta\Bigr) 
X_{,\theta}\nonumber\\
&&+\Sigma P_\mu^2X-\frac{\Sigma}{2}X(X^2-1)\,,  \label{PXeq}\\
&&\nonumber\\
%%%%%%%%%%%%%%% X equation
\frac{X^2}{\beta}P_t&=&
\frac{\triangle}{\Sigma}P_{t,rr}
+\frac{S}{\Sigma}P_{t,\theta\theta}
+\frac{2a\Xi\cot\theta}{\Sigma^3}\Bigl(\rho^2 S -\Delta 
+\frac{a^2}{\ell^2}\Sigma\sin^2\!\theta\Bigr)P_{\phi,\theta}\nonumber\\
&&-\frac{a\Xi}{\Sigma^3}\Bigl(2r(Sa^2\sin^2\!\theta-\Delta) 
+\Sigma\Delta'\Bigr)P_{\phi,r}\nonumber\\
&&+\frac{\cot\theta}{\Sigma^3}\left(S\bigl(\rho^4+a^4\sin^4\!\theta\bigr)
-2a^2\sin^2\!\theta\Bigl(\Delta-\frac{\rho^2\Sigma}{\ell^2}\Bigr)\right)
P_{t,\theta}\nonumber\\
&&-\frac{\sin^2\!\theta}{\Sigma^3}\Bigl(a^2\bigl(2r\rho^2S+\Sigma\Delta'\bigr)
-\frac{2r\rho^2\Delta}{\sin^2\!\theta}\Bigr)P_{t,r}\,,   \label{Pteq}\\
&&\nonumber\\
%%%%%%%%%%%%%%%% P_phi equation
\frac{X^2}{\beta} P_\phi&=&
\frac{\Delta}{\Sigma}P_{\phi,rr}
+\frac{S}{\Sigma}P_{\phi,\theta\theta}
+\frac{\rho^2}{\Sigma^3}\bigl(2r S a^2\sin^2\!\theta +\Sigma \Delta'
-2r\Delta\bigr)P_{\phi,r}\nonumber\\
&&+\frac{\cot\theta}{\Sigma^3}\left( 2a^2\sin^2\!\theta
\Bigl(\Delta-\frac{a^2}{\ell^2}\Sigma \sin^2\!\theta\Bigr)
-S\Bigl(a^2\sin^2\!\theta(\rho^2-\Sigma)+\rho^4\Bigr)\right)
P_{\phi,\theta}\nonumber\\
&&+\frac{2\cot\theta a^3\sin^4\!\theta}{\Xi\Sigma^3}
\left(\Delta-\rho^2\Bigl(1+\frac{r^2}{\ell^2}\Bigr)\right) P_{t,\theta}\nonumber\\
&&+\frac{a\sin^2\!\theta}{\Xi\Sigma^3} \Bigl( 2r
\bigl(\rho^4 S-\Delta(\Sigma+\rho^2)\bigr)
+\rho^2\Sigma\Delta'\Bigr)P_{t,r}\,, \label{Ppeq}
\eea
where $\rho^2=r^2+a^2$ has been introduced for visual clarity,
$\Delta'= d\Delta/dr$, and
\be\label{Psquared}
P_\mu^2=\frac{(\rho^2P_t+a\Xi P_\phi)^2}{\Sigma \Delta}
-\frac{(\Xi P_\phi+a\sin^2\!\theta P_t)^2}{\Sigma S\sin^2\!\theta}\,.
\ee

\subsection{Approximate solution}

As with the original Schwarzschild, Reissner-Nordstrom and Kerr
black holes, it is useful to develop an analytic approximate solution.
Clearly we expect this to make use of the (possibly AdS) Nielsen
Olesen solutions, and to depend on a single function of $r$ and $\theta$.

Consider the function
\be
R\equiv \frac{\rho}{\sqrt{\Xi}}\sin\theta\,,
\ee
which tends to the Kerr expression $\rho\sin\theta$ as $\ell\to\infty$.
Then, assuming that the vortex is much thinner than the black hole
horizon radius means that $\rho$ is always much greater than one, 
and focusing on the core region of the vortex [$R<{\cal O}(10)$] 
means that $\sin\theta\ll1$. We can therefore expand the
metric functions 
\be
\Sigma = \rho^2 \left ( 1 - \frac{a^2 R^2 \Xi}{\rho^4}\right)\simeq \rho^2\,,\quad 
S= \Xi \left ( 1 + \frac{a^2 R^2}{\ell^2 \rho^2}\right)\simeq \Xi\,,
\ee
and derivatives as
\bea
\frac{\partial~}{\partial r} &=& \frac{Rr}{\rho^2} \frac{d~}{dR}\,,\qquad  
\frac{\partial~}{\partial\theta} =\frac{\rho}{\sqrt{\Xi}}
\left ( 1 - \frac{\Xi R^2}{\rho^2} \right)^{1/2} \frac{d~}{dR}
\simeq \frac{\rho}{\sqrt{\Xi}} \frac{d~}{dR}\,, \nonumber \\
\Delta \frac{\partial^2~}{\partial r^2} 
+ S \frac{\partial^2~}{\partial\theta^2}
&=& \left [ S\Bigl( \frac{\rho^2}{\Xi} - R^2 \Bigr) 
+ \frac{\Delta R^2 r^2}{\rho^4} \right] \frac{d^2~}{dR^2}
+ \Bigl( \frac{\Delta a^2}{\rho^4}-S\Bigr) R \frac{d~}{dR} \\
&\simeq& \rho^2 \Bigl( 1 + \frac{\Delta R^2}{\rho^4}\Bigr)
\frac{d^2~}{dR^2}\,, \nonumber
\eea
to leading order in $R/\rho$. This already leads to significant
simplification of several of the terms in \eqref{PXeq}-\eqref{Ppeq}.
Then a little experimentation suggests the following approximate 
functions
\be\label{ApproxSol}
X \simeq X_{0}(R)\,,\quad
P_\phi \simeq P_{0}(R)\,,\quad
P_t \simeq \frac{a}{\rho^2}\Bigl(\frac{\Delta}{\rho^2}
-\Xi\Bigr) P_{0}(R)\,,
\ee
which to leading order give the approximate equations:
\be
\beal
0&= \left ( 1 + \frac{\Delta R^2}{\rho^4}\right ) X_0'' 
+ \left ( 1+ \frac{4R^2}{\ell^2} \right )\frac{X_0'}{R}
- \frac{P_0^2X_0}{R^2}-\frac{X_0}{2}(X_0^2-1)\,,  \\
\frac{X_0^2}{\beta} P_0 &=
\left ( 1 + \frac{\Delta R^2}{\rho^4}\right )P_0'' -
\left ( 1 + \frac{\left( 2\Delta -  r\Delta'\right) R^2}{\rho^4}
\right) \frac{P_0'}{R}\,.
\eeal
\ee
Away from the horizon, $\Delta \sim \rho^4/\ell^2$ to leading
order, and we recover the AdS Nielsen-Olesen equations
\eqref{AdSNO}. However retaining the $R^2/\ell^2$ terms
is perhaps misleading, as we require $\ell>{\cal O}(10)$ 
in order for the horizon radius of an extremal black hole not to be
too small. We also see that on (or near) the horizon, the 
${\cal O}(R^2/\ell^2)$ corrections to the Nielsen-Olesen 
equations fail to have the precise AdS form. This implies that
while we can use the analytic approximation to good effect away
from the black hole, near the horizon we would expect 
corrections to our solution at order ${\cal O}(\ell^{-2})$.

Note that because of the behaviour of $\Delta$ at large $r$,
the approximation for $P_t$ in \eqref{ApproxSol} actually
becomes proportional to $P_\phi$ at large $r$:
$P_t\sim a P_\phi/\ell^2$. Our gauge field is thus
\be
\mathbi{P} = P_\phi d\phi+ P_t dt \sim
P_0(R) \left (d\phi+ \frac{a}{\ell^2} dt \right)\,,
\ee
therefore it would appear that we have an electric field
inside our vortex far from the black hole. In fact, this
is simply an artifact of the Boyer-Lindquist style 
coordinates we have used in \eqref{KNADS}, which
asymptote AdS$_4$ in a rotating frame with angular
momentum $\Omega_\infty=a/\ell^2$ \cite{GPP}.
One may remove this rotation by introducing new variables
\be
\varphi=\phi+\frac{a}{l^2}t\,,\quad T=t\,.
\ee
It is then easy to check that $\mathbi{P}$ in \eqref{ApproxSol} now reads 
\be
\mathbi{P}=P_0(R)\Bigl(d\varphi-\frac{a(2mr-q^2)}{\rho^4}dT\Bigr)\,.
\ee
The $P_T$ component is now negative definite and falls 
off appropriately at large $r$. The form of this solution is
now identical to that used in \cite{GKW}.

\begin{figure}[hb]
\begin{center}
\includegraphics[width=4cm]{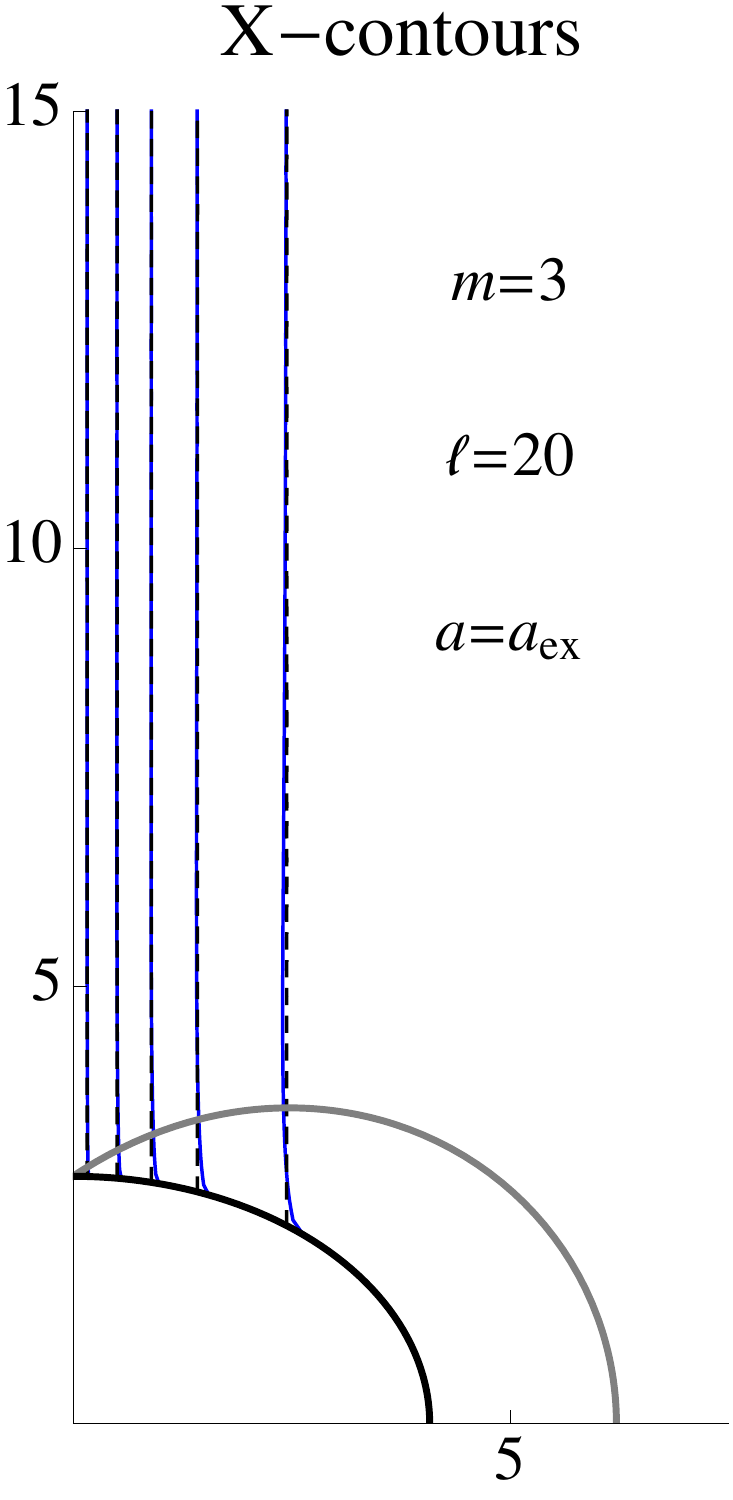}\nobreak
\includegraphics[width=4cm]{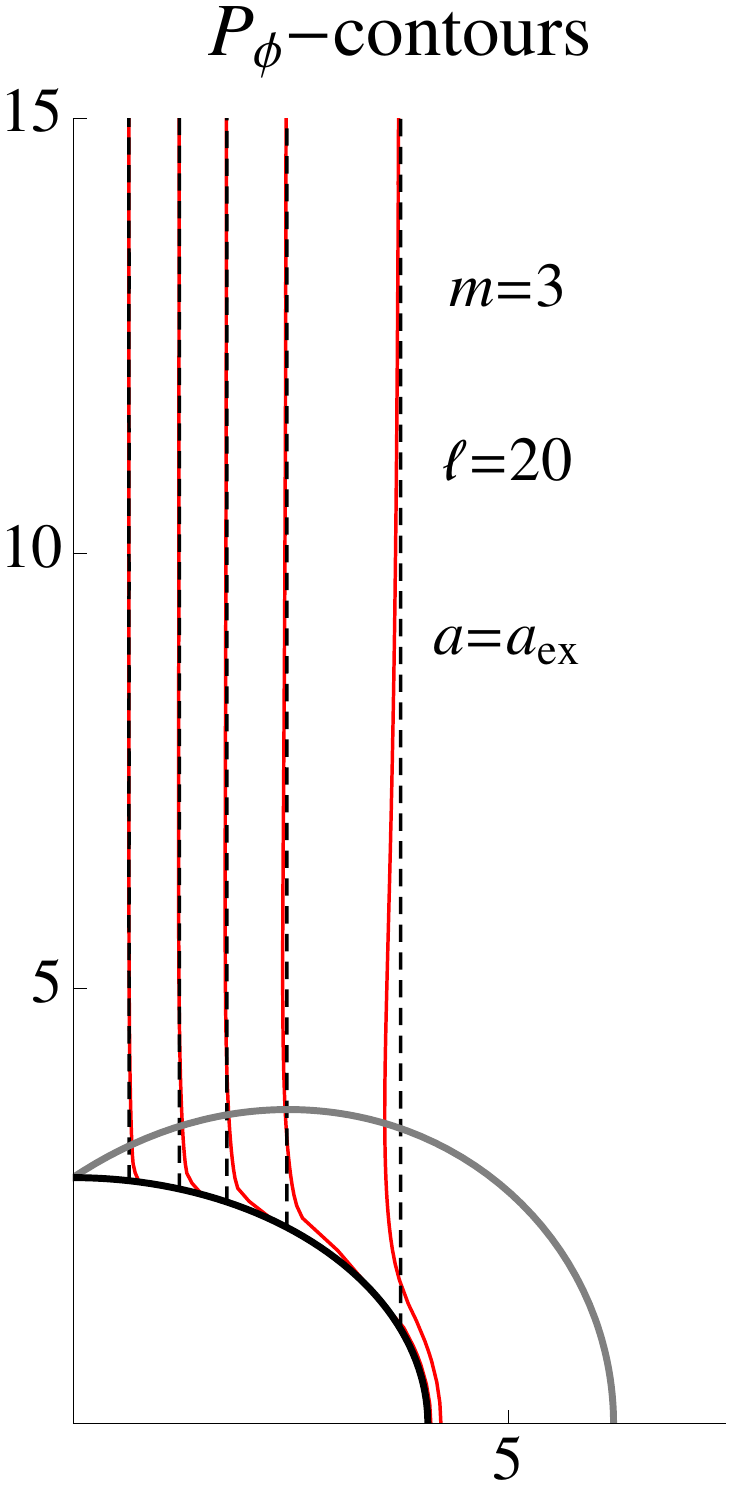}\nobreak
\includegraphics[width=4cm]{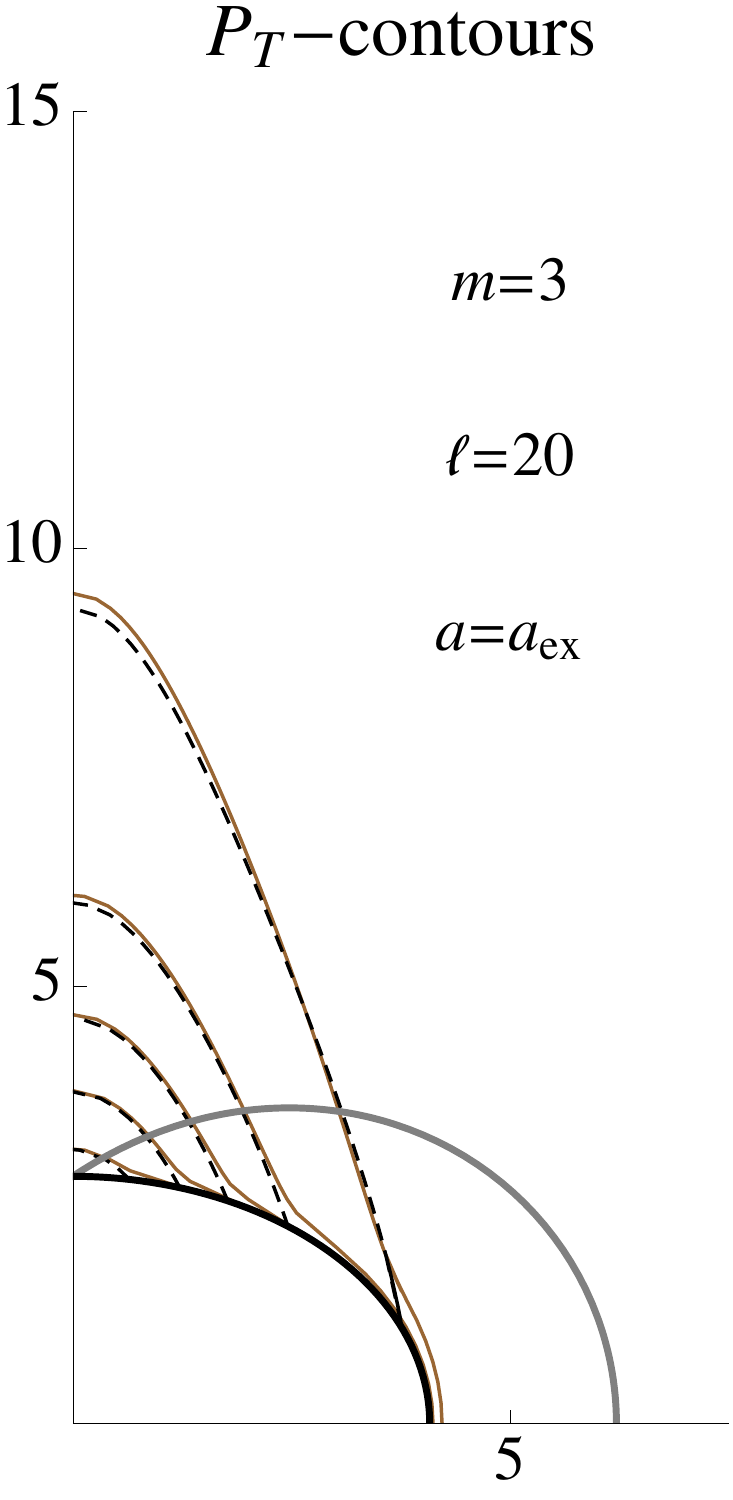}\\
\caption{{\bf Approximate vs. numerical solution:}
In each case the numerical solution
is shown in solid colour, and the approximation in dashed black.
Contours of $0.1-0.9$ (in steps of $0.2$) of the range of each 
field are shown. From left to right: The Higgs field in blue,
the $P_\phi$ field in red, and $P_T$ (the component
with respect to the nonrotating frame at infinity) in brown. 
For $P_T$,
we show contours of $0.1-0.9$ of the maximal negative
value, which is attained on the poles of the horizon.
The outer grey curve represents the boundary of the ergosphere.
}
\end{center}
\label{fig:compare}
\end{figure}

Figure \ref{fig:compare} shows a comparison of this
pseudo-analytic approximation with a numerically obtained
solution for an extremal low mass lowish $\ell$ black hole. 
We take the values $m=3, \ell=20, q=0$, and with $a\simeq2.939$ 
at its extremal value in order to draw a parallel with the plot
in \cite{GKW}.
What is clearly shown is that the approximation is extremely good 
almost everywhere, the only slight discrepancy appearing
near the event horizon -- as expected given the structure of
the corrections to the approximation there.

\subsection{Extremal black holes}

The extremal horizon exhibits 
a Meissner effect for the cosmic string, in which if the black 
hole becomes too `small' the cosmic string magnetic flux
is expelled from the black hole, and the horizon remains in 
the false vacuum.  For both Reissner-Nordstrom \cite{BEG} 
and Kerr \cite{GKW}  black holes,  the existence of this phase 
transition has been proven analytically, as well as
demonstrated  numerically. The Reissner-Nordstrom 
transition is second order, corresponding to a continuous
change in the order parameter (the magnitude of the Higgs
field) between piercing and expelling solutions. For the 
Kerr black hole however, the phase transition was first order, 
corresponding to a discontinuous change in the value
of the gradient of the zeroth component of the gauge field
between piercing and expelling solutions. 

We will now argue for the existence of a Meissner effect in
the AdS-Kerr-Newman black holes; the Kerr-Newman situation 
follows from taking the large-$\ell$ limit. Begin by defining new
variables $P$ and $Q$:
\be
S P = \Xi P_\phi + a \sin^2\!\theta P_t\,, 
\quad (r-r_+)Q = \rho^2 P_t + a \Xi P_\phi\,, 
\ee
where the factors have been chosen so that the
horizon equations are clearly identifiable, and the 
range of $P$ is $P\in[0,1]$.
The field equations \eqref{PXeq}-\eqref{Pteq} become
\bea
0 &=& \frac{\Delta}{\Sigma}X,_{rr}+\frac{\Delta'}{\Sigma}X,_r
+\frac{1}{\Sigma\sin\theta}
\Bigl(S\sin\theta X,_{\theta}\Bigr),_{\theta}
\nonumber \\
&& +\left(\frac{(r-r_+)^2 Q^2}{\Sigma\,\Delta}
-\frac{P^2}{\Sigma\,S\,\sin^2\theta}\right)X -\frac{X}{2}(X^2-1) , 
\label{bulkextX} \\
\frac{X^2P}{\beta} &=& \frac{\Delta}{\Sigma}P,_{rr}
+\frac{S}{\Sigma}P,_{\theta\theta}+\frac{\Sigma\Delta'
-2r\Delta}{\Sigma^2}P_{,r}
+\frac{\cot \theta}{\Sigma}\Bigl(4\frac{a^2}{l^2}\sin^2\!\theta
-\frac{S}{\Sigma}\bigl(\Sigma-2a^2\sin^2\!\theta\bigr)\Bigr)P,_{\theta}
\nonumber \\
&& +\frac{2a \sin^2\theta}{\Sigma^2}
\left((r-r_+)\left (r Q_{,r}-\cot\theta\,Q_{,\theta}- Q\right) 
+aP\Bigl(1-\frac{r^2}{l^2}\Bigr) + rQ\right)\,, 
\label{bulkextP} \\
\frac{X^2 Q}{\beta} &=& \frac{\Delta}{\Sigma}
\frac{\left[(r-r_+)Q\right]_{,rr} }{(r-r_+)}
+\frac{S}{\Sigma}Q_{,\theta\theta}
+ \frac{\cot\theta}{\Sigma^2}(2a^2\sin^2\theta 
(1 + \frac{r^2}{\ell^2}) +S\,\Sigma ) Q_{,\theta}\nonumber \\
&& + \frac{2\Delta}{\Sigma^2} \left (\frac{a}{(r-r_+)} 
(r S P_{,r}-S\cot\theta P_{,\theta}-(2-S)P) -rQ_{,r} -\frac{r_+Q}{(r-r_+)}
\right)\,, \label{bulkextQ} 
\eea
which in the extremal limit and on the horizon reduce to
\bea
\left (S\sin\theta X'\right)'  &=& X \sin\theta \left [
\frac{S P^2}{\sin^2\theta}-\frac{2Q^2}{\Delta''_+}
-\frac{\Sigma_+}{2}(1-X^2) \right] , 
\label{extX} \\
\left ( \frac{S^2 P'}{\Sigma_+\sin\theta}\right)' &=&
P S \sin\theta \left [
\frac{X^2}{\beta \sin^2\theta} 
-\frac{2a^2}{\Sigma_+^2}
\left ( 1 - \frac{r_+^2}{\ell^2}\right) \right]
-\frac{2a r_+S Q \sin\theta}{\Sigma_+^2}
\,, \label{extP} \\
\left ( \frac{S \sin\theta Q'}{\Sigma_+} \right)'
&=&  \frac{X^2 Q}{\beta} \sin\theta 
\,,\label{extQ}
\eea
where a prime now denotes $d/d\theta$, and the ``$+$''
subscript indicates the function is evaluated at $r=r_+$,
given by \eqref{rplusext}. Note that unlike the vacuum 
Kerr case, in which $r_+=a$, there is no simple 
factorization of $\Sigma_+$ leading to a clean
$\theta$-dependence in these equations.

Note that if $a=0$, $Q\equiv0$, and $S\equiv1$ and
 our system of horizon equations reduces {\it precisely}
to the Reissner-Nordstrom horizon equations studied
in \cite{BEG}. Therefore we expect essentially the same
analytic arguments to hold here (which is the case as we shall
see below). Further, since $Q$ vanishes, we expect
a second order phase transition governed by the continuous 
order parameter $X$. On the other hand, if $a\neq0$, $Q$ is
nonzero in the bulk of the spacetime and so we must
examine the full system of horizon equations.

Let us look first at the behaviour of the horizon function $Q$,
as this will give us the order of the phase transition.
For a piercing solution, $X$ is nontrivial on the horizon. 
Hence 
\be
S \beta \sin\theta Q'(\theta) = \Sigma_+ \int_0^\theta
{X^2 Q\sin\theta} d\theta\,, \label{Qcontra}
\ee
upon integrating \eqref{extQ}. 
We can easily see this cannot be true unless
$Q\equiv0$. Evaluating \eqref{Qcontra} at the first point 
at which $Q'=0$ tells us that $\int_0^\theta
{X^2 Q\sin\theta}=0$, but $Q$ is either positive and 
increasing on this range, or negative and decreasing:
in either case, the integrand is positive or negative 
definite, thus cannot be zero.
Therefore $Q\equiv0$ for a piercing solution. On the other
hand, an expelling solution has $X\equiv0$, with $P_\phi=1$,
hence
\be
\label{expelsol}
P=\frac{\Xi \Sigma_+}{\rho_+^2 S}\,, \qquad
Q \equiv - \frac{2ar_+\Xi}{\rho_+^2}\,.
\ee
Given that $Q$ changes in a discontinuous fashion,
we see that the phase transition is first order for nonzero $a$.

\begin{figure}
\begin{center}
\includegraphics[width=0.9\textwidth]{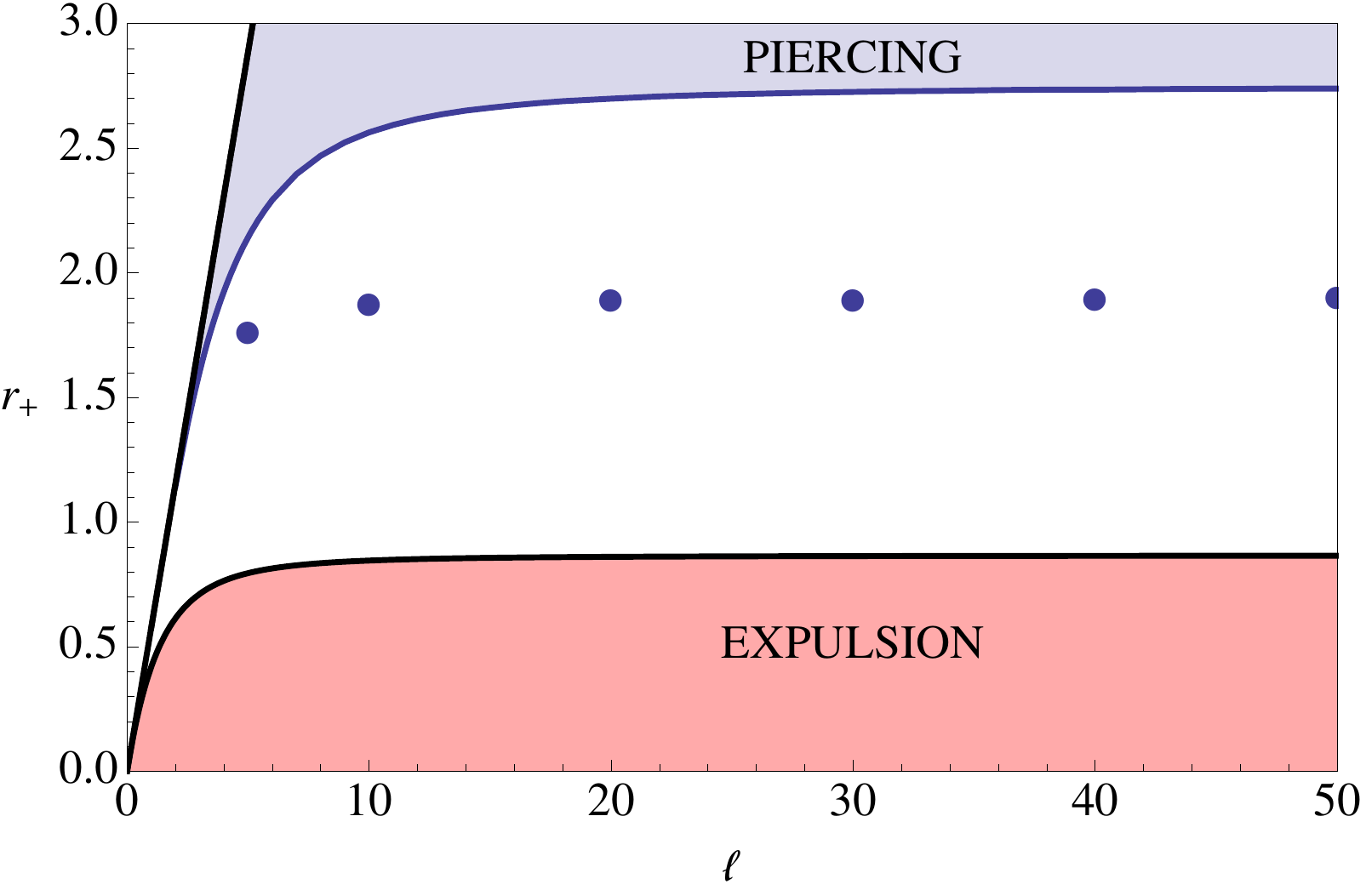}
\caption{{\bf Meissner effect:}
An illustration of the analytic bounds on the critical horizon radius
for the Meissner effect for $q=0$. In the shaded regions, the vortex
should either pierce the horizon, or be expelled as indicated. The
critical radius therefore lies between these two bounds. For sufficiently
low $\ell$, flux is always expelled. Numerically obtained transition 
radii are indicated. The solid $r_+=\ell/\sqrt{3}$ line on the left indicates 
the $a=\ell$ singular limit.}
\label{fig:rcritical}
\end{center}
\end{figure}

It is clear that a flux expelling solution  to the horizon
system of equations \eqref{extX}-\eqref{extQ} can exist. However to 
prove flux expulsion happens, this solution must be extendable to a
bulk solution. To demonstrate this, we follow the argument
of \cite{BEG}. If flux is expelled, $X\equiv0$ on the horizon,
and must become nonzero and positive a small distance from
the horizon, implying $(\Delta X_{,r})_{,r} >0$ just outside
the horizon. Referring to \eqref{bulkextX}, we see therefore 
that
\be
(S \sin\theta X_{,\theta})_{,\theta} + 
\frac{(r_+^2+a^2\cos^2\theta) X}{2}\sin\theta
< \frac{SP^2}{\sin\theta}X <\frac{SX}{\sin\theta}
\ee
is required if a flux expelling solution is to exist. 
Integrating this inequality on $[\theta_0,\pi/2]$ gives
\be
S \sin\theta_0 X_{,\theta_0} > \int_{\theta_0}^{\pi/2} \left (
\frac{(r_+^2+a^2\cos^2\theta)\sin\theta}{2}
- \frac{S}{\sin\theta}\right)Xd\theta  \,.
\ee
Defining $\alpha$ so that $\Sigma_+
\sin^2\alpha/S =2$,  
by taking $\theta_0>\alpha$ we can bound this integral from
below using $X(\theta)>X(\theta_0)$.  We can also
 bound the derivative of $X$ by
$X_{,\theta_0}<\frac{X(\theta_0)-X(\alpha)}{\theta_0-\alpha}
< \frac{X(\theta_0)}{\theta_0-\alpha}$, leading to
\be
S \sin\theta_0 \frac{X(\theta_0)}{\theta_0-\alpha} >
S \sin\theta_0 X_{,\theta_0} 
> X(\theta_0) \int_{\theta_0}^{\pi/2} \left (
\frac{(r_+^2+a^2\cos^2\theta)\sin\theta}{2}
- \frac{S}{\sin\theta}\right) d\theta\,,
\ee
which implies
\be
\frac{({\theta_0-\alpha})}{S(\theta_0)\sin\theta_0} \left (
\frac{r_+^2\cos\theta_0}{2}
+\frac{a^2\cos^3\theta_0}{6}
+\Xi\log\tan\Bigl(\frac{\theta_0}{2}\Bigr)
-\frac{a^2}{l^2}\cos\theta_0
\right)<1
\label{expelinequality}
\ee
on the interval $[\alpha,\pi/2]$. If this inequality is violated,
then we cannot have flux expulsion, and the vortex {\it must}
pierce the black hole. Note, if $a=0$, then \eqref{expelinequality}
is independent of $\ell$, and reduces to the previously 
explored Reissner-Nordstrom
relation  \cite{BEG}, giving the same upper bound on the
horizon radius for flux expulsion of $\sqrt{8.5}$. For $a\neq0$,
we must explore the $\{a,\ell\}$ phase plane (having ensured
that a solution $\alpha$ exists) to determine the upper bound
on the horizon radius. Clearly if $\ell$ drops too low, we require
a large charge to allow for a solution to $\alpha$. Hence for a given
$q$, we expect a minimal value of $\ell$ for this upper bound to exist.
This is shown most clearly for $q=0$, in figure \ref{fig:rcritical}.

To argue that a Meissner effect should exist for sufficiently low 
horizon scales, we assume a piercing solution to
\eqref{extX}-\eqref{extQ} exists, in which 
$X$ and $P$ will have nontrivial profiles symmetric around
$\theta=\pi/2$, with $X$ maximised and $P$ minimised (at
least for large $\ell$ or small $a<q$) at $\pi/2$.
If $a=0$, the argument of \cite{BEG} can be used to
deduce that for $r_+ \lesssim 0.7$ the flux
must be expelled, and this argument can be 
extended to include small $a$ (see appendix).
For $q=0$, or dominant $a$, an alternate argument must
be used. At large $\ell$, $P$ is minimised at $\pi/2$, which
implies a constraint on $r_+$ given by (writing $X_m=X(\pi/2)$):
\be
\beal
P''\left(\frac{\pi}{2}\right) &= P \left ( \frac{X_m^2r_+^2}{\beta} 
- \frac{2a^2}{r_+^2} \Bigl( 1 - \frac{r_+^2}{\ell^2}
\Bigr) \right ) \geq0 \;\;\;
\Rightarrow \;\;\; r_+^4 + 2r_+^2 
\frac{a^2\beta}{\ell^2} > 2 a^2\beta\,.
\eeal
\ee
However, for low values of $\ell$, we cannot show that
$P$ is minimised at $\pi/2$, and indeed scrutiny of piercing
solutions near the phase transition indicates a tiny
modulation in $P$. What we can say however, is that $P$
has at most one additional turning point on $[0,\pi/2]$,
as the source term on the RHS of \eqref{extP} is
monotonically decreasing on $[0,\pi/2]$, hence 
$S^2P'/\Sigma_+\sin\theta$ has at most one turning
point where $X^2 \Sigma_+^2 = 2 a^2\beta \sin^2\theta 
(1-r_+^2/\ell^2)$.

Suppose therefore that we are at low $\ell$ and $P$
has such a turning point on $[0,\pi/2]$. 
Now consider $S^2/\Sigma_+\sin\theta$; the derivative
\be
\left ( \frac{S^2}{\Sigma_+\sin\theta}\right)'
= -\frac{S\cot\theta}{\Sigma_+^2 \sin\theta} \left [
(r_+^2+a^2)\Xi - 3 a^2 \Bigl( 1 + \frac{r_+^2}{\ell^2} \Bigr)\sin^2\theta
+ \frac{a^4}{\ell^2} \sin^4\theta\right]
\ee
 has a zero at $\theta_0$, where
\be
\frac{a^2}{\ell^2} \sin^2\theta_0 = \frac32
\left ( 1 + \frac{r_+^2}{\ell^2} \right) - \frac12
\sqrt{9\left ( 1 + \frac{r_+^2}{\ell^2} \right)^2 - 4 \Xi 
\frac{r_+^2 + a^2}{\ell^2}}\,.
\ee
For $q=0$, $\sin\theta_0\in[0,\sqrt{2/3}]$, as
$\ell$ ranges from $a$ to $\infty$, whereas the node
in $P$ only switches on for lower $\ell$, and initially
appears at $\pi/2$. Therefore at $\theta_0$
we expect $S^2P'/\Sigma_+\sin\theta>0$,
and hence
\be
(r_+^2 + a^2 \cos^2\theta_0)^2 > X^2(\theta_0)
\Sigma_+^2(\theta_0) > 2 a^2\beta \sin^2\theta_0
\Bigl(1-\frac{r_+^2}{\ell^2}\Bigr)\,.
\ee
Thus, if this equality is not satisfied at $\theta_0$, we
deduce that a piercing solution is not possible, and 
expulsion must occur. Figure \ref{fig:rcritical} shows
this lower bound for $q=0$.

The full details of the phase transition must be determined
numerically, and figure \ref{fig:rcritical} shows the
numerically obtained critical horizon radius as a function of
$\ell$ for $q=0$ together with the analytic lower and upper
bounds on $r_{+,crit}$. 
We discuss the phase transition further
in section \ref{discuss}.

\section{Numerical solution}

In order to obtain numerical solutions of the vortex equations 
\eqref{PXeq}-\eqref{Ppeq}, which form an elliptic system, 
we follow references \cite{AGK} and \cite{GKW}, employing 
a gradient flow technique on a two-dimensional polar grid.
Briefly, this method introduces a fictitious time variable,
with the `rate of change' of our functions being proportional to
the actual elliptic equations we wish to solve: 
\be
\dot{Y}^i = \Delta Y^i + F^i (\bf{Y}, \nabla \bf{Y})\,,
\ee
where $\Delta^i$ represents a second order (linear) 
elliptic operator and $F$ is a (possibly nonlinear) function 
of the variables $Y^i$ and their gradients, such that the RHS
is our system of elliptic equations. We now have a 
diffusion problem, and solutions to
this new equation eventually ``relax'' to a steady state, 
in which the variables are no longer changing 
with each time step, and the solutions $Y^i$ satisfy our
elliptic equations. The only subtlety with the given set-up is
that our elliptic system has one boundary (the event horizon)
on which our equations become parabolic. This was 
discussed in detail in \cite{AGK}, with the result that on each
grid update, we update the event horizon, using the horizon
equations, and fixing
\be
P_t=-\frac{a\Xi P_\phi}{r_+^2+a^2}\,
\ee
on the horizon, which is mandated by finiteness of the
energy-momentum tensor.
\begin{figure}
\begin{center}
\includegraphics[width=4.9cm]{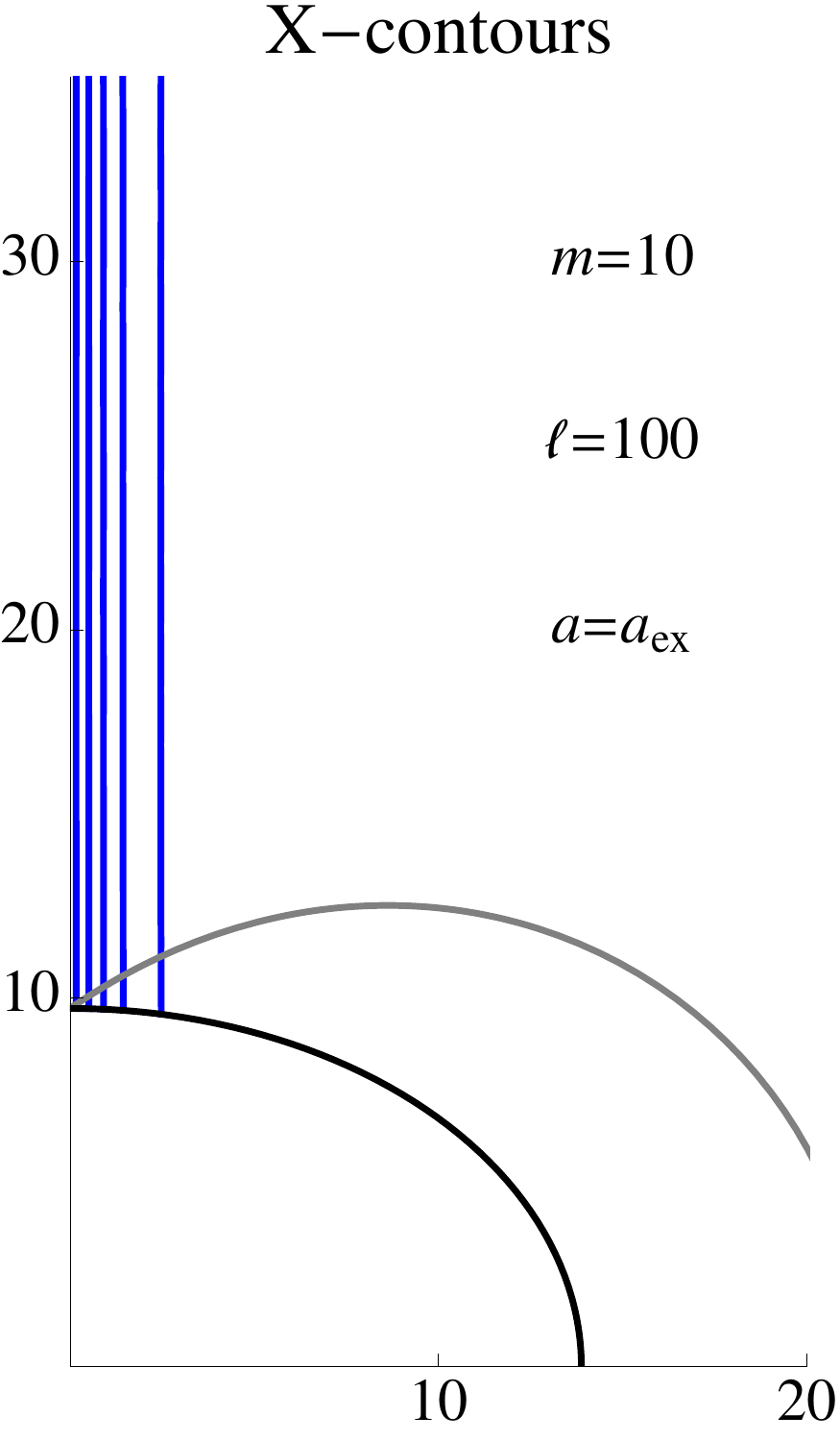}\nobreak
\includegraphics[width=4.9cm]{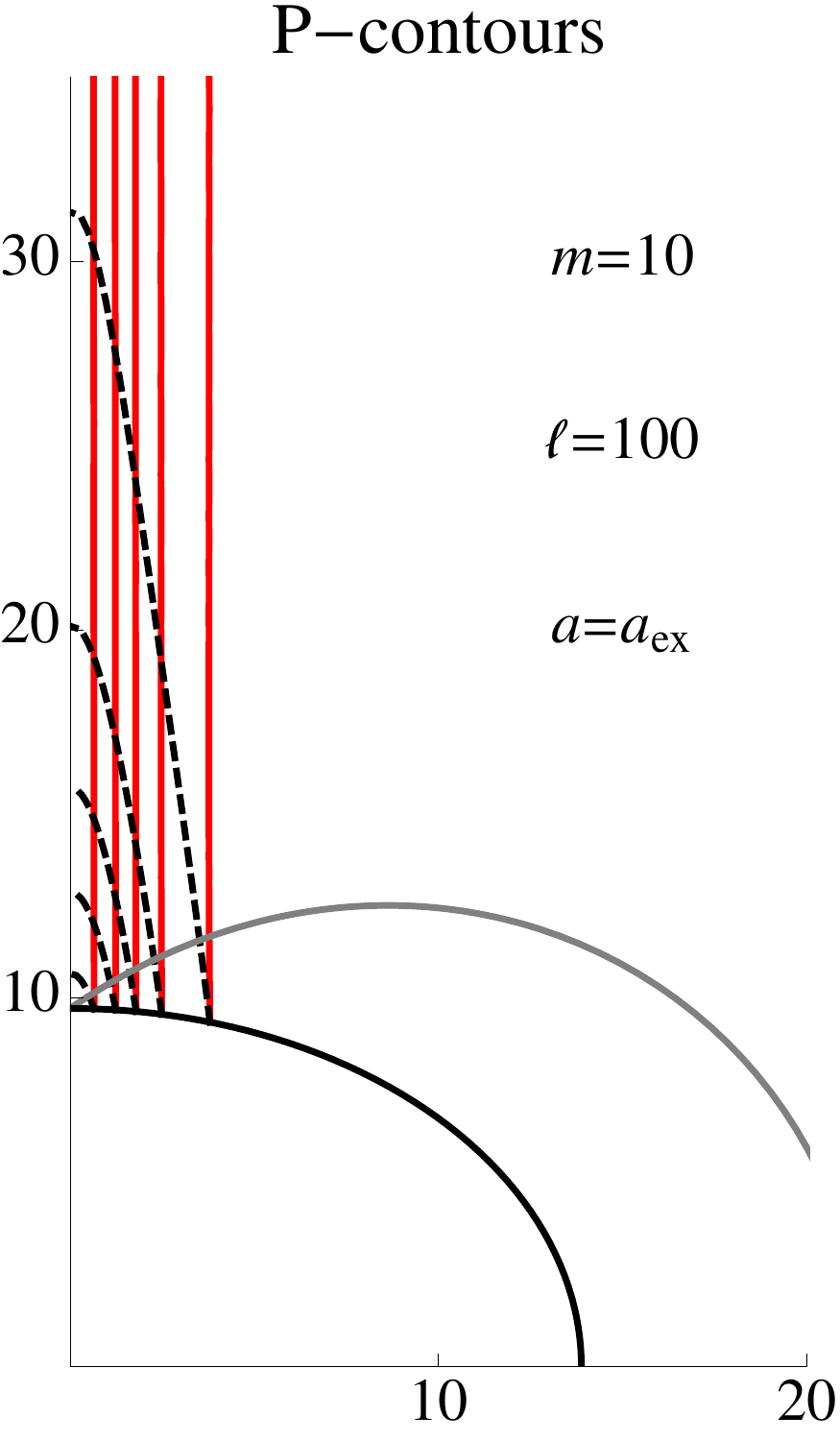}\\
\includegraphics[width=4.9cm]{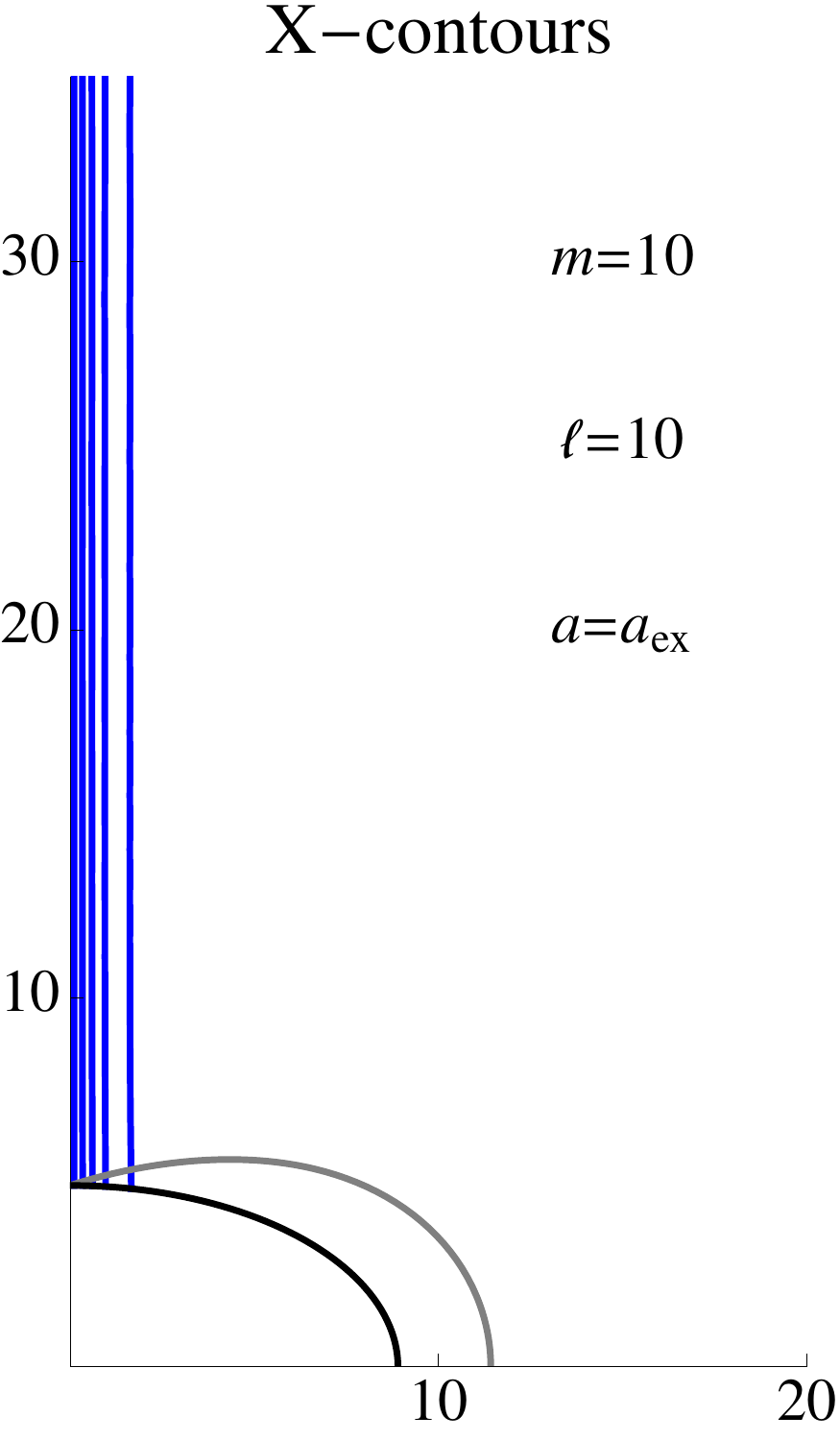}\nobreak
\includegraphics[width=4.9cm]{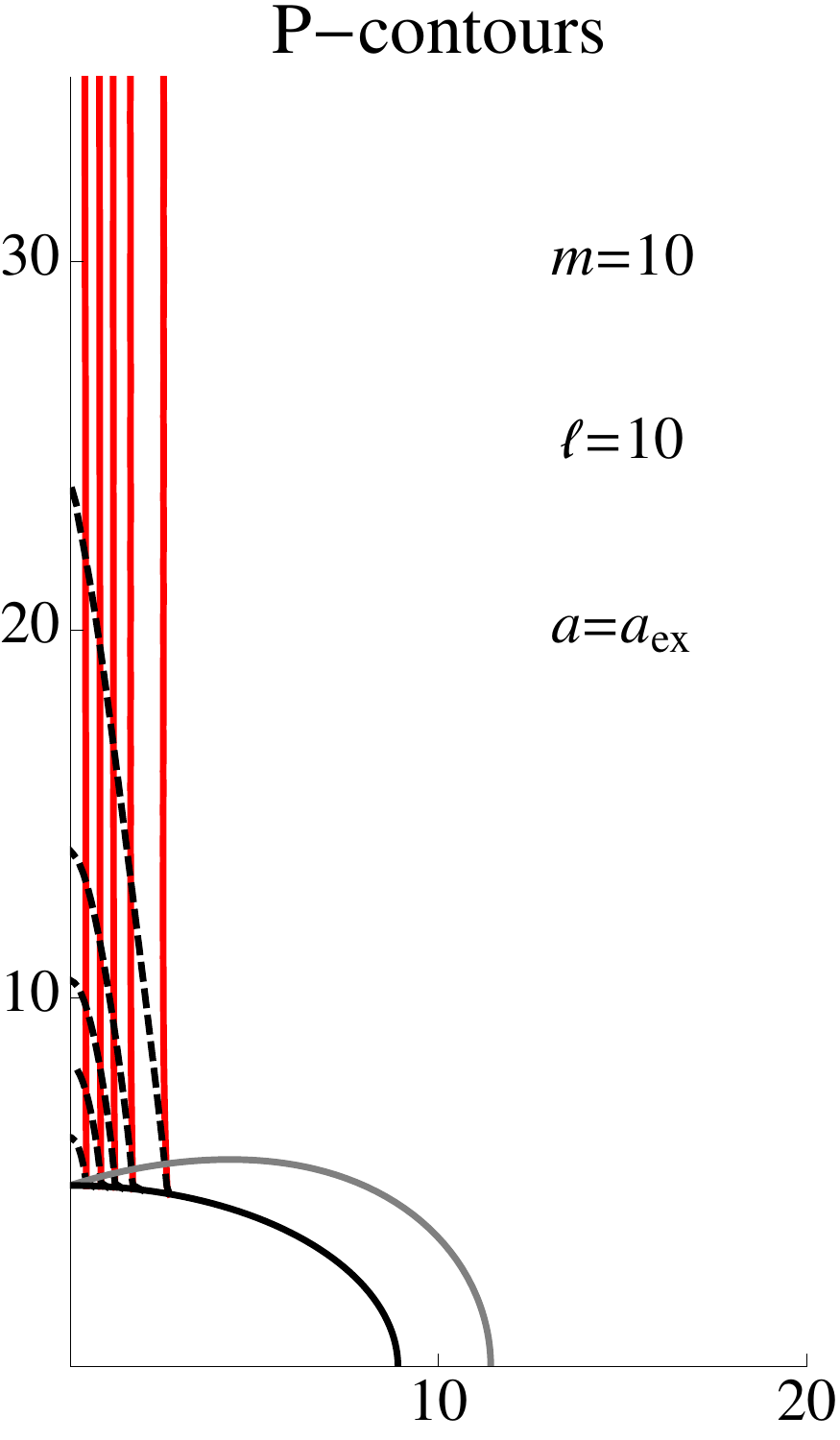}\\
\caption{{\bf AdS-Kerr vortex:} A depiction of the numerical 
solution for the 
AdS-Kerr vortex for an extremal uncharged rotating black hole.
The upper plots have $\ell = 100$, the lower plots
$\ell=10$. In each case, the contours of the
Higgs field are shown on the left in blue ($X=0.1-0.9$ in steps 
of $0.2$), and on the right,
the angular component of the gauge field, $P_\varphi$ 
in red (with the same contour steps as for $X$), and 
$P_T$ in dashed black with contours of $0.1-0.9$ of
$P_{T,min}= -0.0519, -0.116$ for the $\ell=100$ and
$\ell=10$ cases respectively.}
\end{center}
\label{fig:q0}
\end{figure}

As an initial condition for 
the integration, we use the approximate solutions for the 
functions $X$, $P_{\phi}$  and $P_{t}$ given in equations 
\eqref{ApproxSol}, where we obtain the forms for $P_0(R)$ 
and $X_0(R)$ by numerically integrating  \eqref{AdSNO} 
on a one-dimensional grid. The approximate solution is
accurate to order $r^{-2}$, thus we choose our outer
boundary to be sufficiently far from the horizon that our 
analytic approximation is extremely accurate near this
outer radial boundary, which is not updated in our code.
On axis we impose the standard vortex boundary conditions, 
($X=0$, $P_\phi=1$) while leaving $P_t$  to relax 
by continuity. As pointed out in \cite{GKW}, the fact that $P_t$ is 
not restricted can be understood by noting that there is a 
dyonic degree of freedom that is introduced into the solution 
due to the presence of the black hole.
\begin{figure}
\begin{center}
\includegraphics[width=4.9cm]{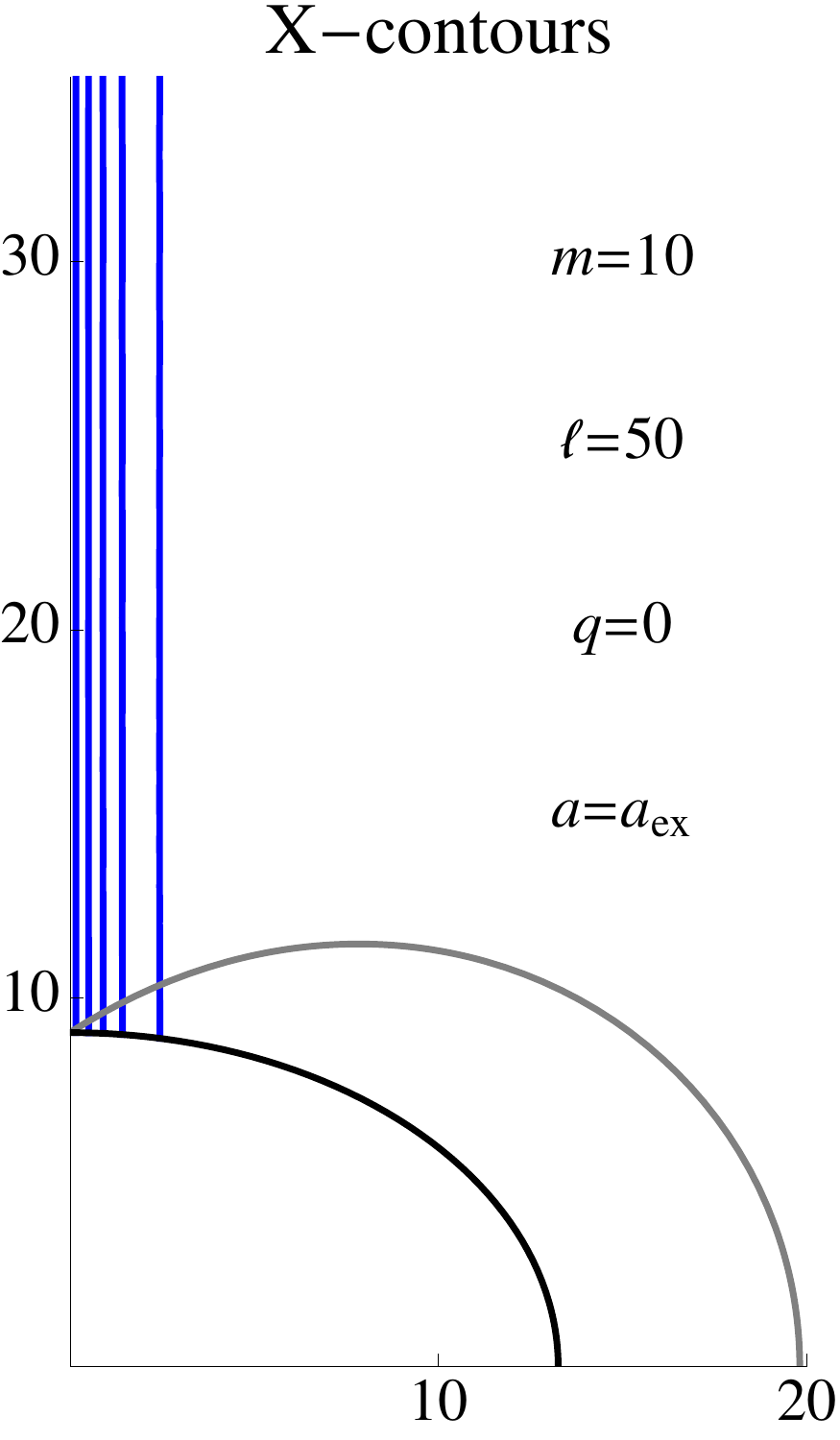}\nobreak
\includegraphics[width=4.9cm]{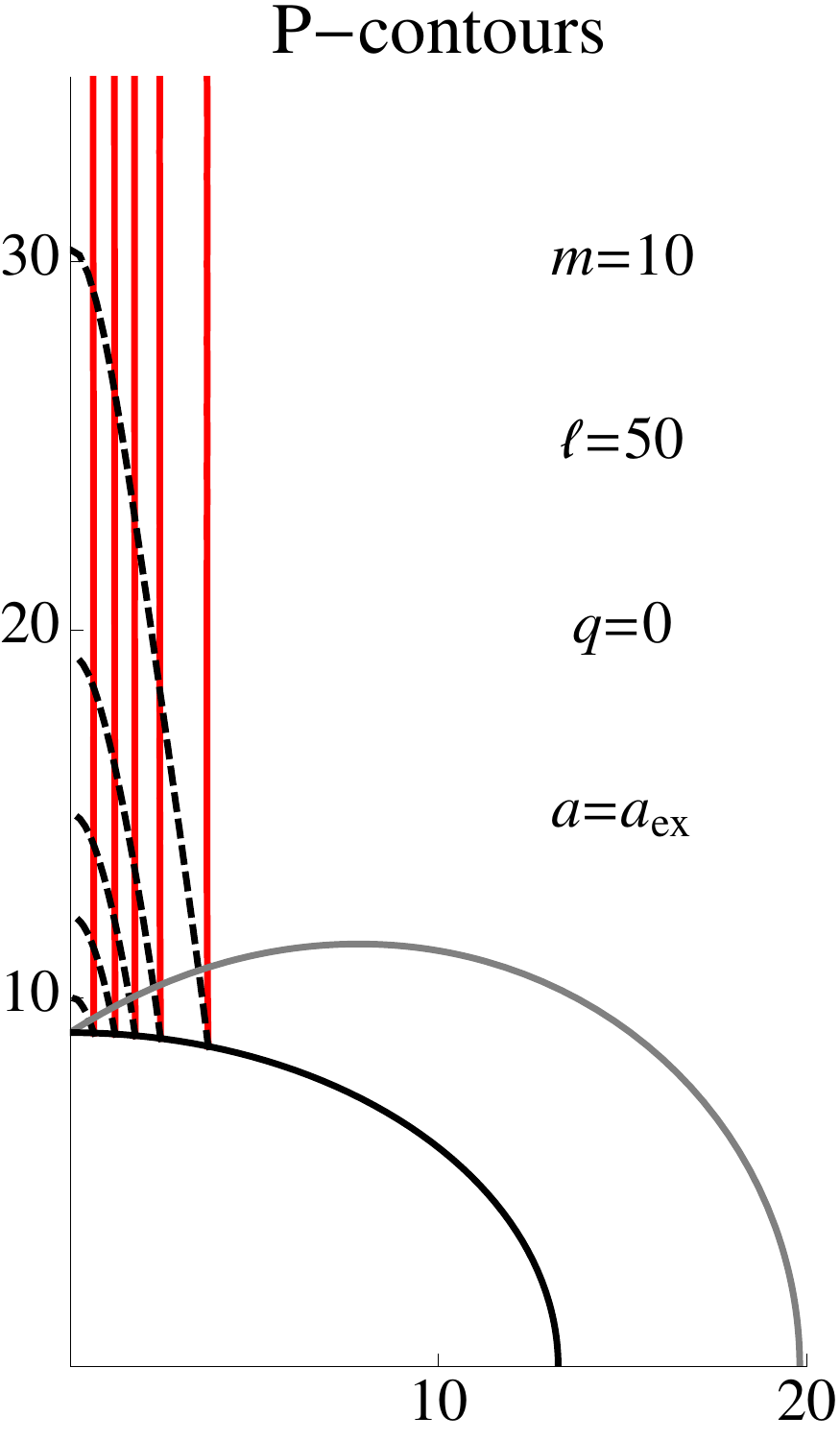}\\
\includegraphics[width=4.9cm]{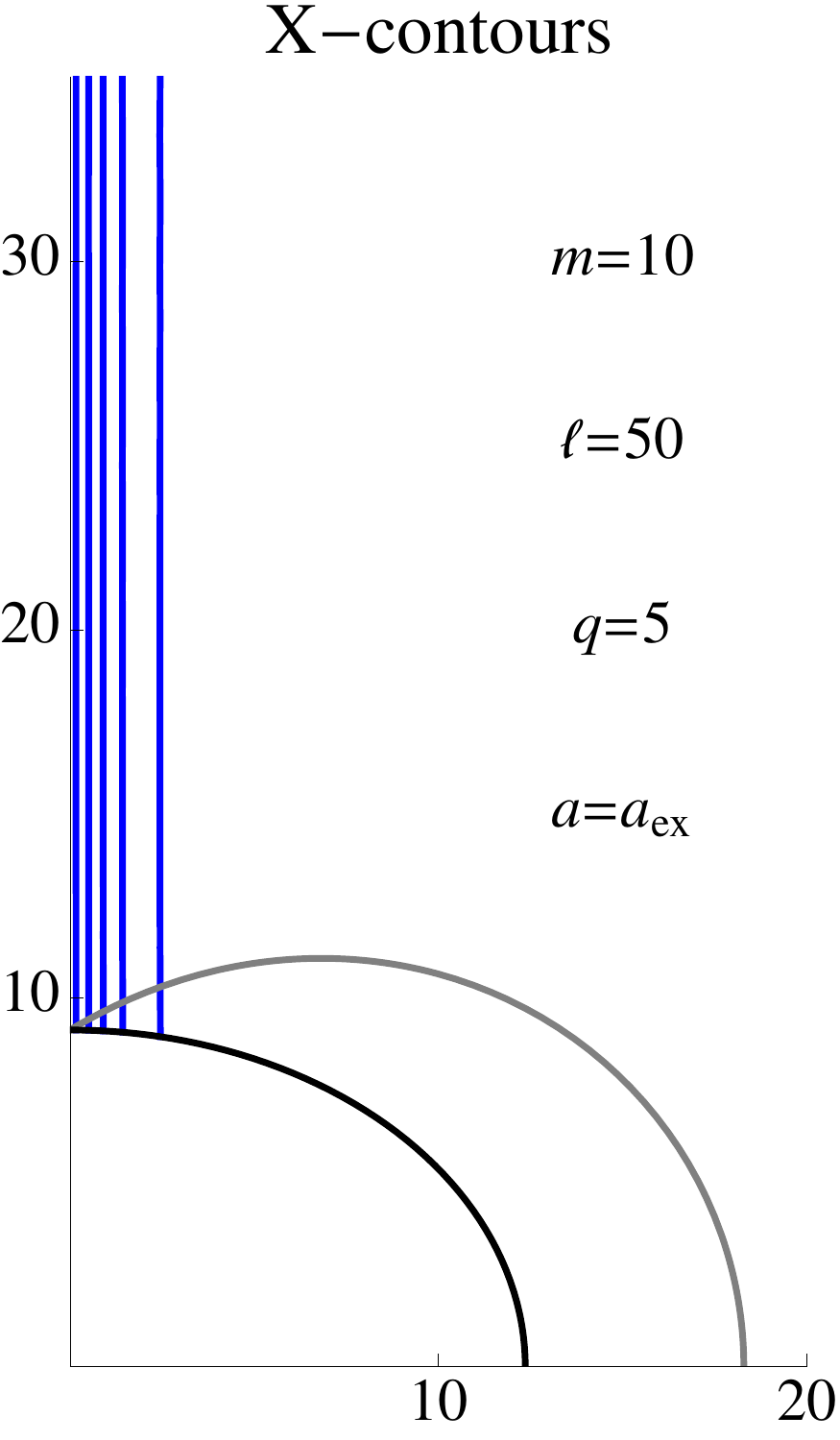}\nobreak
\includegraphics[width=4.9cm]{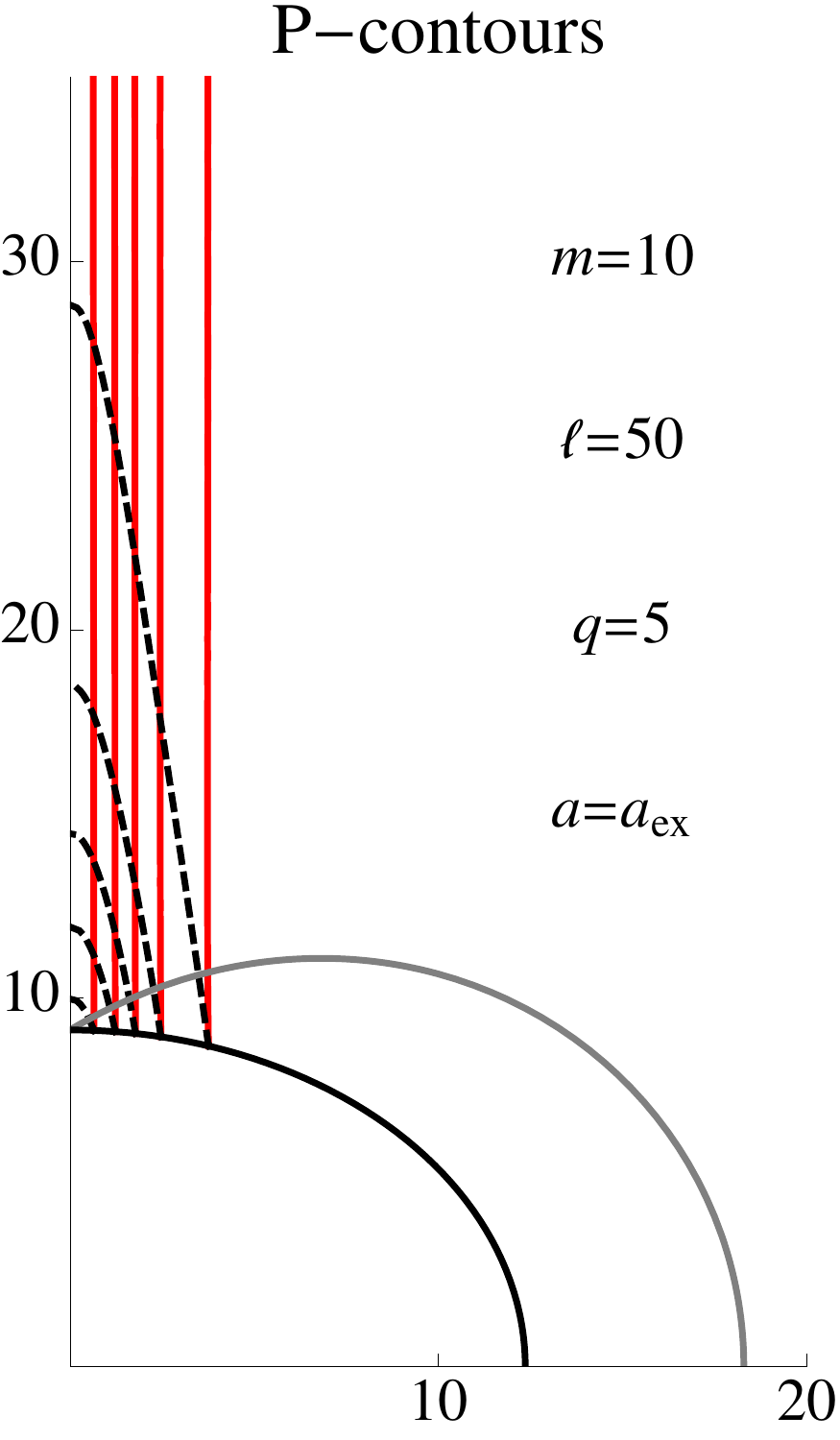}
\caption{{\bf AdS-Kerr-Newman vortex:} Numerical solutions 
for the AdS-Kerr-Newman vortex with 
$\ell = 50$  and $q = 0$, (upper) and $q=5$ (lower) with
the same contour conventions as for figure \ref{fig:q0},
with $P_{T,min} = -0.0569$ for $q=0$, and $P_{T,min}
=-0.0563$ for $q=5$.
}\end{center}
\label{fig:50}
\end{figure}

Figures \ref{fig:q0} and \ref{fig:50} show a selection 
of the solutions obtained 
from the integration method above which highlight the 
effects of the parameters $\ell$ and $q$ on the rotating 
black hole vortex. 
In all plots, we have chosen to illustrate the solution
by plotting contour lines for each field of $0.1-0.9$
of the full range of the field in steps of $0.2$. Thus,
for the $X$ and $P_\phi$ fields, we have shown the
$0.1, 0.3, 0.5, 0.7$, and $0.9$ contours, but for
the $P_T$ field (note -- this is the gauge field component
with respect to a {\it non-rotating} frame at infinity)
the maximally negative value of $P_T$ is attained on
the horizon at the poles. The numerical values of these
contours therefore vary from plot to plot. The 
actual value of $P_{T,min}$ is given in the captions.

Figure \ref{fig:q0} shows the vortex solution for the 
case of $\ell =100$ and $\ell=10$ respectively, at the extremal 
limit with the charge parameter $q$ set to zero. 
The solution away from the extremal limit is similar (see
\cite{GKW}), the main difference being that the actual
numerical values of the $P_T$ contours are lower.
For $\ell=100$, the plots are almost indistinguishable
from the vacuum Kerr vortex solution analysed in \cite{GKW},
however, for $\ell=10$, the effect of the cosmological
constant can be easily seen. Comparing the figures, one notes 
that dropping the value of $\ell$ strongly impacts the size of 
both the black hole horizon as well as the vortex, causing the 
vortex width to tighten, the $P_T$ fields to shrink closer
to the horizon, which itself shrinks significantly. 

Figure \ref{fig:50} then demonstrates the effect of adding a 
non-zero charge to the AdS-Kerr vortex. As can be seen, 
this does not significantly impact the vortex, and appears to 
merely shift the horizon and ergosphere inwards, 
while slightly causing the 
$P_T$ contour lines to creep closer to the horizon, as is
expected since the rotation parameter $a=a_{ex}$ will
be lower with the charged black hole at the same mass.

\section{Discussion}\label{discuss}

We have examined the behaviour and interactions of vortices 
with asymptotically AdS charged and rotating black holes.
We first obtained an approximate solution to the abelian Higgs Model 
in the background of a Kerr-Newman AdS black hole,
and showed that the Nielsen-Olesen equations retain their 
AdS form up to corrections of order $R^2/\ell^2$.  Consequently
we found that our approximation was extremely good  everywhere 
except near the event horizon as expected. The comparison 
illustrated in figure \ref{fig:compare} shows that the actual solution
has a stronger expulsion of flux than the approximation.
Upon transforming to a frame that is non-rotating at the boundary, 
the form of our solution is very close to its asymptotically 
flat counterpart.

\begin{figure}
\begin{center}
\includegraphics[width=0.49\textwidth]{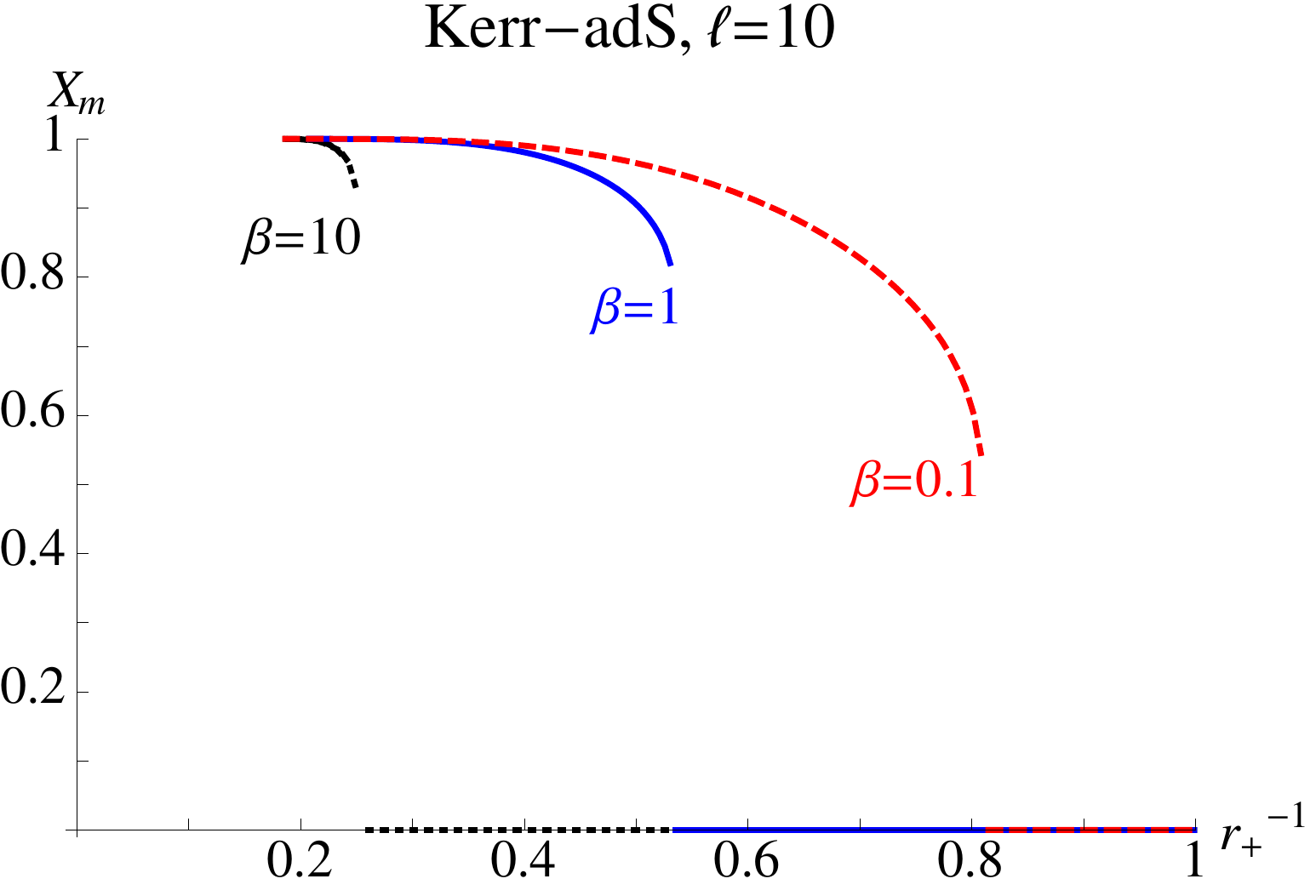}\nobreak
\includegraphics[width=0.49\textwidth]{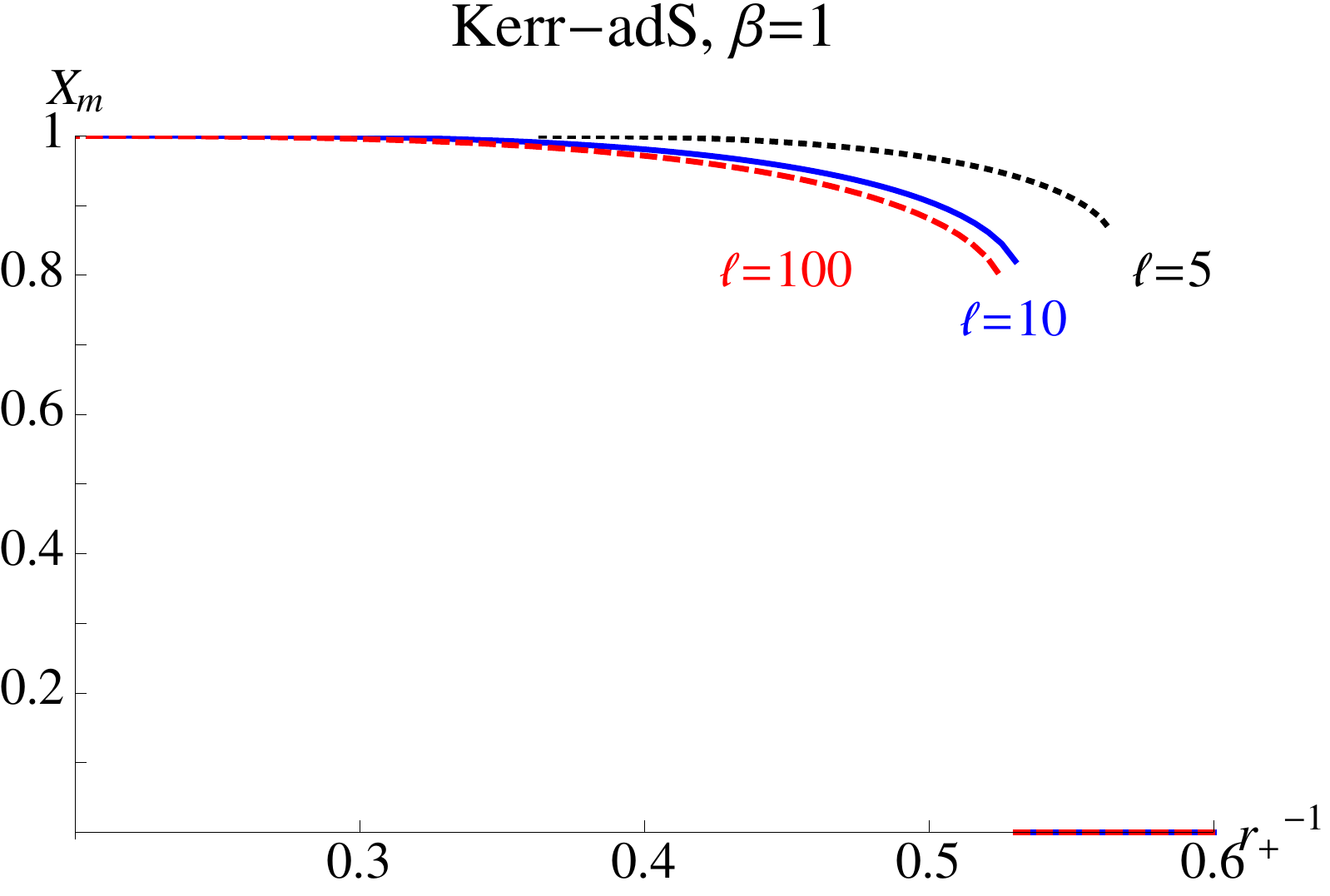}
\caption{{\bf Flux expulsion behavior:}
Plots illustrating features of the flux expulsion phase
transition on the event horizon of the black hole. The
maximal value of the Higgs field $X_m=X(\pi/2)$ is shown
as a function of $r_+^{-1}$ for varying $\beta$ (left) and
$\ell$ (right).
}
\label{fig:phasetrans}
\end{center}
\end{figure}

For extremal black holes we explored the existence of a 
Meissner effect with the cosmic string flux being expelled 
from the black hole at small horizon radii (although one
should be cautious about the stability of such small
black holes \cite{CarDias}). We presented analytic arguments to
show that such a phase transition exists, showing that
in the presence of rotation it is a first order transition.
We numerically explored the phase space to confirm
this expectation, and figure \ref{fig:phasetrans} shows the
numerical results for the phase transition at several
values of $\ell$ and $\beta$. The existence of the first
order transition is confirmed, and the effect of $\ell$ is to 
lower the critical value of $r_+$ at which the transition
occurs. This is also reflected in a drop of both analytic
bounds for expulsion and piercing of the vortex. We also
notice that the value of the order parameter ($X(\pi/2)$)
rises with decreasing $\ell$, seen in the right 
plot of figure \ref{fig:phasetrans}. The left 
plot of figure \ref{fig:phasetrans}
shows the effect of changing $\beta$, and is similar to 
the corresponding plot for the vacuum Kerr solution
in \cite{GKW}. However the effect of changing $\beta$
is far more pronounced at the relatively low value of
$\ell=10$ illustrated. Note that, unlike pure Kerr, the 
plots do not extend to $r_+^{-1}\to0$:  there is an
upper limit on the angular momentum, and hence 
horizon radius. 

The numerical integrations are considerably more sensitive
with the addition of the cosmological constant, mainly because
an additional scale has been added which causes the vortex to 
contract, as well as the black hole. Unfortunately this has prevented 
us from investigating the small-$\ell$ case in significant detail. 
This is  the region of interest for a holographic interpretation of 
our results, though our solution would only be relevant in the IR 
as it does not have the requisite boundary conditions.

Vortices in the bulk  can be interpreted as defects in the dual CFT
\cite{VH,Dias:2013bwa}, where in the IR they are heavy pointlike 
excitations in a superfluid around  which the phase of the 
condensate winds. A vortex must have a core radius since the 
vanishing of the condensate at its location is energetically costly and 
so must happen over some finite region. A recent study 
\cite{Dias:2013bwa} of vortices with planar black holes has
indicated that  their IR physics can be understood from the 
viewpoint of a defect or boundary CFT \cite{Cardy:2004hm}.  
A study of holographic superconductivity in the context of 
(topologically spherical) rotating black holes \cite{Sonner} 
found that the superconducting state in the dual theory  
(for certain choices of parameters) can be destroyed for 
sufficiently large rotation. The localization of the condensate 
depends on the sign of the mass-squared term of the scalar, 
with a droplet/ring-like structure appearing for positive/negative 
values of this term.  The instability towards forming vortex 
anti-vortex pairs depends on this sign \cite{Sonner}.

It would be interesting to study these effects further in light of 
our results.  Even without rotation we obtain a Meissner effect, 
and so very small black holes with expelled flux can exist.  
Their interpretation in the context of the boundary theory (as well 
as distinguishing them from the flux-pierced case) remains to be 
understood, perhaps in
terms of the absence of a mass gap for the flux-expelled case.

\acknowledgments

RG is supported in part by STFC (Consolidated Grant ST/J000426/1),
in part by the Wolfson Foundation and Royal Society, and in part
by Perimeter Institute for Theoretical Physics. DK is supported by 
Perimeter Institute. DW is supported by an STFC studentship. DW
would also like to thank Perimeter Institute for hospitality.
Research at Perimeter Institute is supported by the Government of
Canada through Industry Canada and by the Province of Ontario through the
Ministry of Research and Innovation. This work was supported in part
by the Natural Sciences and Engineering Research Council of Canada.

\appendix
\section{Lower bound argument}
\label{app:bounds}

Following \cite{BEG},
assume a piercing solution exists, then \eqref{extX} and \eqref{extP}
have smooth solutions for $X$ and $P$ in which $X$ increases
from zero at the poles to a maximum, $X_m$ at the equator,
and $P$ decreases from $1$ at the poles to a minimum, $P_m$,
at the equator.
Evaluating \eqref{extX} and \eqref{extP} at the equator gives
the relations:
\bea
X''\left ( \frac{\pi}{2} \right) &= X_m \left [ P^2_m 
+ \frac{r_+^2}{2} \left ( X^2_m-1\right)\right] \leq 0 
\ &\Rightarrow \ P_m^2 \leq \frac{r_+^2}{2} \left ( 1- X^2_m
\right) \leq \frac{r_+^2}{2} \label{Xdpbd}\,,\\
P''\left ( \frac{\pi}{2} \right) &= P_m \left [ \frac{X_m^2
r_+^2}{\beta} - \frac{2a^2}{r_+^2} \left ( 1 
- \frac{r_+^2}{\ell^2}\right) \right] \geq 0 \
&\Rightarrow \ r_+^4\geq X_m^2 r_+^4 \geq 2 a^2 
\beta\Bigl( 1 - \frac{r_+^2}{\ell^2}\Bigr)\,.\quad
\label{Pdpbd}
\eea
Since $P\leq1$, the first relation gives no new information unless
$r_+<\sqrt{2}$, so we will assume this from now on. The second
relation clearly gives no information if $a=0$; however, for nonzero
$a$ and sufficiently small $q$, the bound \eqref{Pdpbd} is violated
at sufficiently low $\ell$, or indeed if $q<a\lesssim0.6$ for all $\ell$.

If $a$ is sufficiently small that \eqref{Pdpbd} does not give useful 
information, then we can
bound $r_+$ by a generalisation of the argument in \cite{BEG}.
Assuming a piercing solution, \eqref{Pdpbd} bounds $P''(\pi/2)$
above by:
\be
P''\left ( \frac{\pi}{2} \right) \leq \frac{r_+}{\sqrt{2}} \left [ \frac{X_m^2
r_+^2}{\beta} - \frac{2a^2}{r_+^2} \Bigl( 1 
- \frac{r_+^2}{\ell^2}\Bigr)\right] \sqrt{1-X_m^2}
\leq \frac{\sqrt{2}r_+^3}{3\sqrt{3} \beta} \left ( 1 - \frac{2a^2\beta}
{r_+^4} \Bigl( 1 - \frac{r_+^2}{\ell^2}\Bigr)\right)^{3/2}\!\!\!,
\ee
where we use \eqref{Xdpbd}, and maximise over $X_m$ in the second
inequality.

\begin{figure}
\begin{center}
\includegraphics[width=0.9\textwidth]{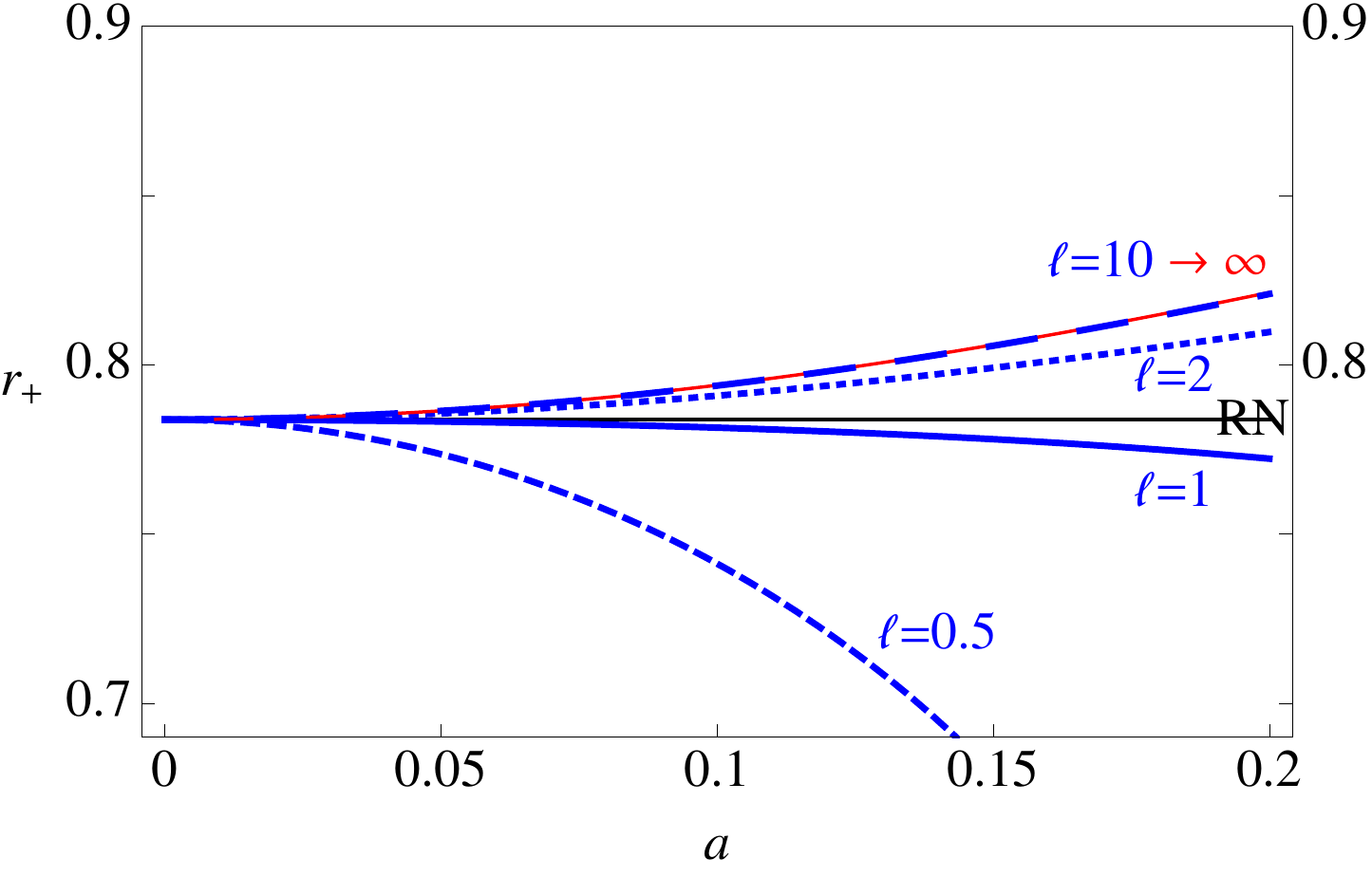}
\caption{{\bf Expulsion bound:} An illustration of the running 
of the lower bound with $a$.
The bound is plotted for $\ell=0.5,1,2,10,\infty$ as labeled. For
$\ell>5$, the curve changes very little, as can be seen by the
infinite $\ell$ curve depicted by a thin red line. The value at
$a=0$ is the RN value obtained in \cite{BEG}, and is shown 
as the horizontal solid black line.
}
\end{center}
\label{fig:lowerrun}
\end{figure}

To get a lower bound on $P''$ we use $P''(\pi/2)\geq -P'(\theta_0)
/(\pi/2-\theta_0)$, where $\theta_0$ is where $P'$ is maximally
negative, \eqref{extP} giving:
\be
P'\left(\theta_0\right) = -\frac{P \tan\theta}{\beta}
\frac{\Sigma_+^2 X^2 - 2 a^2 \beta \sin^2\theta
\left ( 1 - r_+^2/\ell^2\right)}{S(\Sigma_+-2a^2\sin^2\theta)
-4\Sigma_+ (a^2/\ell^2) \sin^2\theta}
\Bigg| _{\theta=\theta_0}\,.
\ee
Thus
\be
\beal
\left |P'\left(\theta_0\right) \right| &\leq
\frac{P(\theta_0) \tan\theta_0}{\beta}
\frac{\Sigma_+^2(\theta_0) - 2 a^2 \beta \sin^2\theta_0
\left ( 1 - r_+^2/\ell^2\right)}
{S(\theta_0)(\Sigma_+(\theta_0)-2a^2\sin^2\theta_0)
-4\Sigma_+(\theta_0) (a^2/\ell^2) \sin^2\theta_0}\\
&\leq \frac{\bigl(r_+^4-2a^2\beta \left ( 1 - r_+^2/\ell^2\right)\bigr) 
\,\tan\theta_0}{\bigl(r_+^2(1-4a^2/\ell^2)-2a^2\bigr)\,\beta}\,.
\eeal
\ee
Clearly for this bound to be meaningful, we also require 
$r_+^2(1-4a^2/\ell^2)>2a^2$, so we will assume this
going forward. We therefore have that
\be
\frac\pi2-\theta_0<
\cot\theta_0 \leq 
\frac{r_+^4-2a^2\beta \left ( 1 - r_+^2/\ell^2\right) }
{\bigl(r_+^2(1-4a^2/\ell^2)-2a^2\bigr)\,\beta\,|P_0'|}\,,
\ee
meanwhile
\be
1-\frac\pi2|P_0'| < P_m<\frac{r_+}{\sqrt{2}} \qquad\Rightarrow\qquad
|P_0'| > \frac2\pi \Bigl( 1 - \frac{r_+}{\sqrt{2}}\Bigr)\,,
\ee
giving
\be
P''\left (\frac\pi2\right) \geq \frac{|P_0'|}{\pi/2-\theta_0}\geq
\beta |P_0'|^2 \frac{r_+^2(1-4a^2/\ell^2)-2a^2}
{r_+^4-2a^2\beta \left ( 1 - r_+^2/\ell^2\right) }\,.
\ee
Folding this in with the upper bound on $P''(\pi/2)$, we
see that for a piercing solution to exist the following inequality
must hold:
\be
\frac{6\sqrt{6}\beta^2}{\pi^2}
\Bigl( 1 - \frac{r_+}{\sqrt{2}} \Bigr)^2
\left (1-\frac{2a^2\beta}{r_+^4} 
\Bigl( 1 - \frac{r_+^2}{\ell^2} \Bigr)\right)^{-5/2} 
\frac{r_+^2(1-4a^2/\ell^2)-2a^2}{r_+^7} <1\,,
\label{lowerrn}
\ee
with $2>r_+^2>2a^2/(1-4a^2/\ell^2)$ and $r_+^4
+2 a^2 r_+^2/\ell^2 >2a^2\beta$. The former of these bounds places a 
stronger constraint on $a$, but in fact the constraint \eqref{lowerrn}
breaks down before even this is violated. Since the value of $r_+$ 
satisfying \eqref{lowerrn} is quite low (just less than one), the relation
gives no useful information once $a$ gives a significant contribution
to $r_+$. For large $\ell$, this happens around $a\sim0.7$, but for 
$\ell$ of order unity or below, this happens at a much lower value 
($\sim0.3$ for $\ell=1$). We illustrate the running of this lower bound
with $a$ in figure \ref{fig:lowerrun}.

The actual value of this bound is less important than the fact it exists,
which then implies the existence of a phase transition on the event
horizon and flux expulsion.

\vskip 5mm

\providecommand{\href}[2]{#2}
\begingroup\raggedright\endgroup


\begin{thebibliography}{10}

\bibitem{nohair}
%\bibitem{Chrusciel:1994sn}
P.~T.~Chrusciel,
{\it `No hair' theorems: Folklore, conjectures, results,}
Contemp.\ Math.\  {\bf 170}, 23 (1994).
[gr-qc/9402032].

\bibitem{ChaseAdler}
J.~E.~Chase,
{\it Event Horizons in Static Scalar-Vacuum Space-Times,}
Com.\ Math.\ Phys.\ {\bf 19} 276--288 (1970).\\
%\cite{Adler:1978dp}
S.~L.~Adler and R.~B.~Pearson,
{\it `No Hair' Theorems for the Abelian Higgs and Goldstone Models,}
Phys.\ Rev.\ {\bf D18}, 2798 (1978).

\bibitem{nohair2}
%\cite{Bekenstein:1996pn}
J.~D.~Bekenstein,
{\it Black hole hair: 25 - years after,}
In {\em Moscow 1996, 2nd International A.D. Sakharov
Conference on physics} 216-219
[\href{http://xxx.lanl.gov/abs/gr-qc/9605059}{{\tt gr-qc/9605059}}].

\bibitem{otherhair}
%\cite{Bizon:1990sr}
P.~Bizon,
{\it Colored black holes,}
Phys.\ Rev.\ Lett.\  {\bf 64}, 2844 (1990).\\
%\cite{Luckock:1986tr}
H.~Luckock and I.~Moss,
{\it Black Holes Have Skyrmion Hair,}
Phys.\ Lett.\ {\bf B176}, 341 (1986).\\
%\cite{Lee:1991vy}
K.~-M.~Lee, V.~P.~Nair and E.~J.~Weinberg,
{\it Black holes in magnetic monopoles,}
Phys.\ Rev.\ {\bf D45}, 2751 (1992)
[\href{http://xxx.lanl.gov/abs/hep-th/9112008}{{\tt hep-th/9112008}}].

\bibitem{coshair}
%\cite{Jacobson:1999vr}
T.~Jacobson,
{\it Primordial black hole evolution in tensor scalar cosmology,}
Phys.\ Rev.\ Lett.\  {\bf 83}, 2699 (1999)
[\href{http://xxx.lanl.gov/abs/astro-ph/9905303}{{\tt astro-ph/9905303}}].\\
%\cite{Chadburn:2013mta}
S.~Chadburn and R.~Gregory,
{\it Time dependent black holes and scalar hair,}
\href{http://xxx.lanl.gov/abs/1304.6287}{{\tt [arXiv:1304.6287 [gr-qc]]}}.\\
%\cite{Abdalla:2013ara}
E.~Abdalla, N.~Afshordi, M.~Fontanini, D.~C.~Guariento and E.~Papantonopoulos,
{\it Cosmological black holes from self-gravitating fields,}
\href{http://xxx.lanl.gov/abs/1312.3682}{{\tt [arXiv:1312.3682 [gr-qc]]}}.

\bibitem{Kerrscalar}
%\cite{Hod:2012px}
S.~Hod,
{\it Stationary Scalar Clouds Around Rotating Black Holes,}
{\em Phys.\ Rev.\ D} {\bf 86}, 104026 (2012)
[Erratum-ibid.\ D {\bf 86}, 129902 (2012)]
\href{http://xxx.lanl.gov/abs/1211.3202}{{\tt [arXiv:1211.3202 [gr-qc]]}}.\\
%\cite{Herdeiro:2014goa}
C.~A.~R.~Herdeiro and E.~Radu,
{\it Kerr black holes with scalar hair,}
{\em Phys.\ Rev.\ Lett.\ } {\bf 112}, 221101 (2014)
[\href{http://arXiv.org/abs/arXiv:1403.2757}{{\tt arXiv:1403.2757 [gr-qc]}}].

\bibitem{Vilenkin}
%\cite{Vilenkin:1984ib}
A.~Vilenkin,
{\it Cosmic Strings and Domain Walls,}
Phys.\ Rept.\  {\bf 121}, 263 (1985).\\
A.~Vilenkin and E.~P.~S. Shellard,
{\em Cosmic Strings and other Topological Defects}.
\newblock Cambridge University Press, Cambridge, England, 1994.

\bibitem{IpS}
%\cite{Ipser:1983db}
J.~Ipser and P.~Sikivie,
{\it The Gravitationally Repulsive Domain Wall,}
Phys.\ Rev.\ {\bf D30}, 712 (1984).\\
%\cite{Gibbons:1993in}
G.~W.~Gibbons,
{\it Global structure of supergravity domain wall space-times,}
Nucl.\ Phys.\ {\bf B394}, 3 (1993).

\bibitem{Vilenkin2}
%\cite{Vilenkin:1981zs}
A.~Vilenkin,
{\it Gravitational Field of Vacuum Domain Walls and Strings,}
Phys.\ Rev.\ {\bf D23}, 852 (1981).

\bibitem{EGS}
%\bibitem{Emparan:2000fn}
R.~Emparan, R.~Gregory and C.~Santos,
{\it Black holes on thick branes,}
Phys.\ Rev.\ {\bf D63}, 104022 (2001)
[\href{http://xxx.lanl.gov/abs/hep-th/0012100}{{\tt hep-th/0012100}}].
  
\bibitem{AGK}
%\cite{Achucarro:1995nu}
A.~Achucarro, R.~Gregory, and K.~Kuijken,
{\it {Abelian Higgs hair for black holes}},
Phys.\ Rev.\  {\bf D52} (1995) 5729--5742,
[\href{http://xxx.lanl.gov/abs/gr-qc/9505039}{{\tt gr-qc/9505039}}].

\bibitem{KinW}
%\cite{Kinnersley:1970zw}
W.~Kinnersley and M.~Walker,
{\it Uniformly accelerating charged mass in general relativity,}
Phys.\ Rev.\ {\bf D2}, 1359 (1970).

\bibitem{AFV}
%\cite{Aryal:1986sz}
M.~Aryal, L.~Ford, and A.~Vilenkin,
{\it {Cosmic strings and black holes}},
Phys.\ Rev.\  {\bf D34} (1986) 2263.

\bibitem{RGMH}
%\cite{Gregory:1995hd}
R.~Gregory and M.~Hindmarsh,
{\it {Smooth metrics for snapping strings}},
Phys.\ Rev.\  {\bf D52} (1995) 5598--5605,
[\href{http://xxx.lanl.gov/abs/gr-qc/9506054}{{\tt gr-qc/9506054}}].\\
%\cite{Eardley:1995au}
D.~M.~Eardley, G.~T.~Horowitz, D.~A.~Kastor and J.~H.~Traschen,
{\it Breaking cosmic strings without monopoles,}
Phys.\ Rev.\ Lett.\  {\bf 75}, 3390 (1995)
[\href{http://xxx.lanl.gov/abs/gr-qc/9506041}{{\tt gr-qc/9506041}}].\\
%\cite{Hawking:1995zn}
S.~W.~Hawking and S.~F.~Ross,
{\it Pair production of black holes on cosmic strings,}
Phys.\ Rev.\ Lett.\  {\bf 75}, 3382 (1995)
[\href{http://xxx.lanl.gov/abs/gr-qc/9506020}{{\tt gr-qc/9506020}}].\\
%\cite{Emparan:1995je}
R.~Emparan,
{\it Pair creation of black holes joined by cosmic strings,}
Phys.\ Rev.\ Lett.\  {\bf 75}, 3386 (1995)
[\href{http://xxx.lanl.gov/abs/gr-qc/9506025}{{\tt gr-qc/9506025}}].\\
%\cite{Achucarro:1997bd}
A.~Achucarro and R.~Gregory,
{\it Selection rules for splitting strings,}
Phys.\ Rev.\ Lett.\  {\bf 79}, 1972 (1997)
[\href{http://xxx.lanl.gov/abs/hep-th/9705001}{{\tt hep-th/9705001}}].

\bibitem{CCESa}
%\cite{Chamblin:1997gk}
A.~Chamblin, J.~Ashbourn-Chamblin, R.~Emparan, and A.~Sornborger,
{\it {Can extreme black holes have (long) Abelian Higgs hair?}},
Phys.\ Rev.\  {\bf D58} (1998) 124014,
[\href{http://xxx.lanl.gov/abs/gr-qc/9706004}{{\tt gr-qc/9706004}}].\\
%\bibitem{CCESb}
%\cite{Chamblin:1997bh}
A.~Chamblin, J.~Ashbourn-Chamblin, R.~Emparan, and A.~Sornborger,
{\it {Abelian Higgs hair for extreme black holes and selection rules for
snapping strings}},
Phys.\ Rev.\ Lett.\  {\bf 80} (1998) 4378--4381,
[\href{http://xxx.lanl.gov/abs/gr-qc/9706032}{{\tt gr-qc/9706032}}].\\
%\bibitem{BG}
%\cite{Bonjour:1998cm}
F.~Bonjour and R.~Gregory,
{\it {Comment on `Abelian Higgs hair for extremal black holes and
selection rules for snapping strings'}},
Phys.\ Rev.\ Lett.\  {\bf 81} (1998) 5034,
[\href{http://xxx.lanl.gov/abs/hep-th/9809029}{{\tt hep-th/9809029}}].

\bibitem{BEG}
%\cite{Bonjour:1998rf}
F.~Bonjour, R.~Emparan, and R.~Gregory,
{\it {Vortices and extreme black holes: The Question of flux expulsion}},
Phys.\ Rev.\  {\bf D59} (1999) 084022,
[\href{http://xxx.lanl.gov/abs/gr-qc/9810061}{{\tt gr-qc/9810061}}].

\bibitem{ROG}
%\cite{Moderski:1998me}
R.~Moderski and M.~Rogatko,
{\it{Abelian Higgs hair for electrically charged dilaton black holes,}}
Phys.\ Rev.\  {\bf D58}, 124016 (1998)
[\href{http://xxx.lanl.gov/abs/hep-th/9808110}{{\tt hep-th/9808110}}].\\
L.~Nakonieczny and M.~Rogatko,
{\it Abelian-Higgs hair on stationary axisymmetric black hole 
in Einstein-Maxwell-axion-dilaton gravity},
Phys.~Rev.~ {\bf D88}, 084039 (2013)
[\href{http://arXiv.org/abs/arXiv:1310.5929}{{\tt arXiv:1310.5929 [hep-th]}}].

\bibitem{GB}
%\cite{Ghezelbash:2001pq}
A.~Ghezelbash and R.~B. Mann,
{\it {Abelian Higgs hair for rotating and charged black holes}},
Phys.\ Rev.\  {\bf D65} (2002) 124022,
[\href{http://xxx.lanl.gov/abs/hep-th/0110001}{{\tt hep-th/0110001}}].

\bibitem{GKW}
%\cite{Gregory:2013xca}
R.~Gregory, D.~Kubiznak and D.~Wills,
{\it Rotating black hole hair,}
JHEP {\bf 1306}, 023 (2013)
\href{http://xxx.lanl.gov/abs/1303.0519}{{\tt [arXiv:1303.0519 [gr-qc]]}}.

\bibitem{GMdS}
%\cite{Ghezelbash:2002cc}
A.~Ghezelbash and R.~Mann, {\it {Vortices in de Sitter space-times}},
{\em Phys.Lett.} {\bf B537} (2002) 329--339,
[\href{http://xxx.lanl.gov/abs/hep-th/0203003}{{\tt hep-th/0203003}}].

\bibitem{VH}
%\cite{oai:arXiv.org:hep-th/0105134}
M.~H.~Dehghani, A.~M.~Ghezelbash and R.~B.~Mann,
{\it Vortex holography,}
Nucl.\ Phys.\ {\bf B625}, 389 (2002)
[\href{http://xxx.lanl.gov/abs/hep-th/0105134}{{\tt hep-th/0105134}}].

\bibitem{DGM1}
%\bibitem{DGM2}
%\cite{Dehghani:2001nz}
M.~Dehghani, A.~Ghezelbash, and R.~B. Mann, 
{\it {Abelian Higgs hair for AdS-Schwarzschild black hole}},  
Phys.\ Rev.\  {\bf D65} (2002) 044010,
[\href{http://xxx.lanl.gov/abs/hep-th/0107224}{{\tt hep-th/0107224}}].

\bibitem{Wald}
%\cite{Wald:1974np}
R.~Wald,
{\it {Black hole in a uniform magnetic field}},
Phys.\ Rev.\  {\bf D10} (1974) 1680--1685.
  
\bibitem{Expulsion1}
A.~R.~King, J.~P.~Lasota, and W.~Kundt, 
{\it Black holes and magnetic fields},
Phys. Rev. {\bf D12} 3037 (1975).

\bibitem{Expulsion2} J.~Bi\v{c}\'ak and L. Dvo\v{r}\'ak,  
{\it Stationary electromagnetic fields around black holes. 
II. General solutions and the fields of some special sources 
near a Kerr black hole} 
General Relativity and Gravitation, {\bf 7} 959 (1976)\\
J.~Bi\v{c}\'ak and L. Dvo\v{r}\'ak, 
{\it Stationary electromagnetic fields around black holes. III. 
General solutions and the fields of current loops near the 
Reissner-Nordstr\"om black hole},
Phys. Rev. {\bf D22} 2933 (1980).

\bibitem{Expulsion3}
A.~Chamblin, R.~Emparan and G.~W.~Gibbons,
{\it Superconducting p-branes and extremal black holes},
Phys.~Rev.~{\bf D58}, 084009 (1998), 
[\href{http://arXiv.org/abs/arXiv:9806017}{{\tt arXiv:9806017 [hep-th]}}].

\bibitem{Penna}
R.~F.~Penna, {\it The Black Hole Meissner Effect and Blandford-Znajek Jets},
[\href{http://arXiv.org/abs/arXiv:1403.0938}{{\tt arXiv:1403.0938 [astro-ph.HE]}}].

\bibitem{Dias:2013bwa} 
O.~J.~C.~Dias, G.~T.~Horowitz, N.~Iqbal and J.~E.~Santos,
{\it Vortices in holographic superfluids and superconductors as conformal defects}
[\href{http://arXiv.org/abs/arXiv:1311.3673} {{\tt arXiv:1311.3673 [hep-th]}}].

\bibitem{Bog}
%\cite{Bogomolny:1975de}
E.~B. Bogomolnyi,
{\it The stability of classical solutions},
Sov.\ J.\ Nucl.\ Phys. {\bf 24} (1976) 449.

\bibitem{NO}
%\cite{Nielsen:1973cs}
H.~B. Nielsen and P.~Olesen,
{\it {Vortex Line Models for Dual Strings}},
Nucl.\ Phys.\  {\bf B61} (1973) 45--61.

\bibitem{BF}
%\cite{Breitenlohner:1982jf}
P.~Breitenlohner and D.~Z.~Freedman,
{\it Stability in Gauged Extended Supergravity,}
Annals Phys.\  {\bf 144}, 249 (1982).

\bibitem{CarterCMP}
%\cite{Carter:1968ks}
B.~Carter,
{\it Hamilton-Jacobi and Schrodinger separable solutions of 
Einstein's equations,}
Commun.\ Math.\ Phys.\  {\bf 10}, 280 (1968).

%\cite{Gibbons:2004ai}
\bibitem{GPP} 
G.~W.~Gibbons, M.~J.~Perry and C.~N.~Pope,
{\it The First law of thermodynamics for Kerr-anti-de Sitter black holes,}
Class.\ Quant.\ Grav.\  {\bf 22}, 1503 (2005)
[\href{http://xxx.lanl.gov/abs/hep-th/0408217}{{\tt hep-th/0408217}}].

\bibitem{CarDias}
%\cite{Cardoso:2004hs}
V.~Cardoso and O.~J.~C.~Dias,
{\it Small Kerr-anti-de Sitter black holes are unstable,}
{\em Phys.\ Rev.\ D} {\bf 70}, 084011 (2004)
[\href{http://xxx.lanl.gov/abs/hep-th/0405006}{{\tt hep-th/0405006}}].

\bibitem{Cardy:2004hm} 
J.~L.~Cardy, {\it Boundary conformal field theory}
[\href{http://arXiv.org/abs/arXiv:hep-th/0411189} {{\tt arXiv:hep-th/0411189}}].

%\cite{Sonner:2009fk}
\bibitem{Sonner} 
J.~Sonner,
{\it A Rotating Holographic Superconductor,}
Phys.\ Rev.\  {\bf D80}, 084031 (2009)
[\href{http://arXiv.org/abs/arXiv:0903.0627}{{\tt arXiv:0903.0627 [hep-th]}}].

\end{thebibliography}
\end{document}